\documentclass[a4paper,11pt]{article}

\usepackage{jheppub} 

\usepackage{hyperref}

\usepackage{booktabs}
\usepackage{xspace}

\usepackage[T1]{fontenc} 

\definecolor{darkgreen}{rgb}{0,0.5,0}
\definecolor{darkpurple}{rgb}{0,0.5,0.5}
\definecolor{darkblue}{rgb}{0,0,0.7}
\definecolor{darkred}{rgb}{0.5,0,0.0}
\definecolor{darkorange}{rgb}{0.8,0.4,0.0}
\definecolor{green}{rgb}{0.0,0.8,0.4}

\newcommand{\as}{\alpha_s}

\newcommand{\GeV}{\;\mathrm{GeV}}
\newcommand{\TeV}{\;\mathrm{TeV}}
\newcommand{\ptjv}{p_{\rm t,veto}}

\newcommand{\LLR}{\text{LL$_R$}\xspace}
\newcommand{\NNLL}{\text{NNLL}\xspace}
\newcommand{\NNLLNNLO}{\text{NNLO+NNLL}\xspace}
\newcommand{\NNLLNNNLO}{N$^3$LO+NNLL\xspace}
\newcommand{\NNLO}{\text{NNLO}\xspace}
\newcommand{\NNNLO}{\text{N$^3$LO}\xspace}
\newcommand{\NNNLL}{\text{N$^3$LL}\xspace}

\newcommand{\sigmai}[1]{{\sigma^{(#1)}}}
\newcommand{\sigmatot}[1]{\sigma_{\rm tot, #1}}

\title{Jet-vetoed Higgs cross section in gluon fusion at N$^3$LO+NNLL
  with small-$R$ resummation}

\author[1]{Andrea Banfi,}
\author[2]{Fabrizio Caola,}
\author[2,3,4]{Fr\'ed\'eric A. Dreyer,}
\author[5]{Pier F. Monni,}
\author[2,*]{Gavin P. Salam\note[*]{On leave from CNRS, UMR 7589, LPTHE, F-75005, Paris, France},}
\author[2,5]{Giulia Zanderighi,}
\author[6]{Falko Dulat}

\affiliation[1]{Department of Physics and Astronomy, University of
  Sussex, Brighton BN1 9RH, UK} \affiliation[2]{CERN, PH-TH, CH-1211
  Geneva 23, Switzerland} \affiliation[3]{Sorbonne Universit\'es, UPMC
  Univ Paris 06, UMR 7589, LPTHE, F-75005, Paris, France}
\affiliation[4]{CNRS, UMR 7589, LPTHE, F-75005, Paris, France}
\affiliation[5]{Rudolf Peierls Centre for Theoretical
  Physics,University of Oxford, Oxford OX1 3NP, UK}
\affiliation[6]{Institute for Theoretical Physics, ETH Z\"urich, 8093
  Z\"urich, Switzerland}

\abstract{We present new results for the jet-veto efficiency and
  zero-jet cross section in Higgs production through gluon fusion.
  We incorporate the N$^3$LO corrections to the total cross section,
  the NNLO corrections to the 1-jet rate, NNLL resummation for the
  jet $p_t$ and LL resummation for the jet radius dependence.
  Our results include known finite-mass corrections and are obtained
  using the jet-veto efficiency method, updated relative to earlier
  work to take into account what has been learnt from the new
  precision calculations that we include.
  For 13 TeV collisions and using our default choice for the
  renormalisation and factorisation scales, $\mu_0=m_H/2$, the matched
  prediction for the jet-veto efficiency increases the pure \NNNLO{}
  prediction by about 2\% and the two have comparable
  uncertainties. Relative to NNLO+NNLL results, the new prediction is
  2\% smaller and the uncertainty reduces from more than 10\% to less
  than 5\%.  Results are also presented for the central scale $\mu_0=m_H$.}

\preprint{CERN-PH-TH-2015-261; OUTP-15-29P}

\begin{document} 
\maketitle
\flushbottom

\section{Introduction}

Since the announcement of the discovery of the Higgs
boson~\cite{Aad:2012tfa,Chatrchyan:2012xdj}, a dynamic research
programme has come into place to measure and constrain its properties.
The precision of the measurements is already such that the
interpretation of data is sometimes limited by theoretical
uncertainties (see
e.g. Refs.~\cite{ATLAS:2014aga,Chatrchyan:2013iaa}).
Experimental errors will decrease in Run II, because of
the higher luminosity and of the higher energy. Experimental analyses
will also benefit from the experience gained in Run I, which will
result in optimised Higgs analyses already from the early stages of
Run II.

The dominant Higgs-production mode at the LHC is gluon-gluon
fusion. 
The most fundamental quantity is the total Higgs production
cross-section, which allows one to compute the total number of Higgs
bosons produced at the LHC for a given energy and luminosity.  
In some Higgs boson decay modes (most notably $WW^*$ and $\tau\tau$), it
is standard to perform different analyses depending on the number of
accompanying jets.
%
%
This is because different jet multiplicities have different dominant
backgrounds.
Of particular importance for the $WW$ decay is the zero-jet case, where
the dominant top-quark decay background is dramatically reduced.
For precision studies it is important to predict accurately the
fraction of signal events that survive the zero-jet constraint, and to
assess the associated theory uncertainty.
Jet-veto transverse momentum thresholds used by ATLAS and CMS are
relatively soft ($\sim 25-30$ GeV), hence QCD real radiation is
severely constrained by the cut and the imbalance between virtual and
real corrections results in logarithms of the form $\ln(\ptjv/m_H)$
that should be resummed to all orders in the coupling constant. 
This resummation has been carried out to next-to-next-to-leading
logarithmic accuracy (\NNLL, i.e.\ including all
terms $\as^n \ln^k (\ptjv/m_H)$ with $k \ge n-1$ in the logarithm of the
cross section) and matched to next-to-next-to-leading order (\NNLO) in
Refs.~\cite{Banfi:2012jm,Stewart:2013faa,Becher:2013xia}
(some of the calculations also included partial \NNNLL{} contributions).
At this order one finds that the effect of the resummation is to shift
central predictions only moderately, and to reduce somewhat the
theoretical uncertainties.
Yet, the residual theoretical uncertainty remains sizeable, roughly
$10\%$~\cite{Banfi:2012jm}, and the impact of higher-order effects
could therefore be significant.

Since the first \NNLLNNLO{} predictions for the jet-veto, three
important theoretical advances happened: firstly, the \NNNLO{}
calculation of the total gluon-fusion cross
section~\cite{Anastasiou:2015ema}; secondly the calculation of the
\NNLO{} corrections to the Higgs plus one-jet
cross-section~\cite{Boughezal:2015dra,Boughezal:2015aha,Caola:2015wna}; and finally
the LL resummation of logarithms of the jet-radius
$R$~\cite{Dasgupta:2014yra}.
Given these recent advances, we are now in a position to improve on
the previous prediction by extending the matching of the jet-veto
cross-section to N$^3$LO+NNLL+\LLR. In order to perform the matching
and to estimate the uncertainties one needs to extend the matching
schemes introduced in Ref.~\cite{Banfi:2012jm} to one order higher.
In doing so, we will also revisit the formulation of the ``jet-veto
efficiency'' (JVE) approach that was introduced in Ref.~\cite{Banfi:2012yh}.

For accurate predictions it is also important to investigate the
impact of finite quark masses, a subject extensively discussed in the
literature.
Finite quark-mass effects are known
exactly only up to
NLO~\cite{Spira:1995rr,Spira:1997dg,Harlander:2005rq,Anastasiou:2006hc,Aglietti:2006tp,Bonciani:2007ex}. The
impact of top quark effects on the leading jet's transverse momentum
at NLO has been studied through a $1/m_t$
expansion~\cite{Neumann:2014nha}.  Different prescriptions have been
proposed to include top and bottom effects in analytic
resummations~\cite{Mantler:2012bj,Grazzini:2013mca,Banfi:2013eda} as
well as parton-shower simulations, e.g. in (N)NLO+PS
generators~\cite{Bagnaschi:2011tu,Bagnaschi:2015qta,Hamilton:2015nsa}.
Here, we include exact mass effects up to NNLL+NLO and study the
impact of the resummation scale associated with the bottom and
top-bottom-interference contributions. Mass effects at \NNLO{} and
\NNNLO{} are currently unknown, so we use the large-$m_t$ limit
(without any rescaling) at these orders.

This paper is organised as follows. 
In Sec.~\ref{sec:JVE} we recall the Jet Veto Efficiency (JVE) method,
and we give a new prescription for the uncertainty estimate. This
differs from the one given in ref.~\cite{Banfi:2012jm}, and we believe
is more appropriate now that the Higgs total production and Higgs plus one-jet 
cross sections are known respectively known through \NNNLO{} and NNLO.
In the rest of Sec.~\ref{sec:ingredients}, we
introduce the various ingredients of the calculation, and we discuss
how they are combined together.
In Sec.~\ref{sec:all} we present our new results at 13 TeV
centre-of-mass energy, while Sec.~\ref{sec:conclu} contains our
conclusions. In Appendix~\ref{app:schemes} we further motivate the
introduction of the new JVE uncertainty prescription. In
Appendix~\ref{app:mho2vsmh} we compare our final predictions obtained
with central scale $m_H/2$ to predictions obtained with central scale
$m_H$. Finally, in Appendix~\ref{app:smallr} we give some technical
details about the small-$R$ resummation.

\section{Outline of the formalism}
\label{sec:ingredients}

\subsection{Updated jet-veto efficiency method at fixed order}
\label{sec:JVE}

The core element of our estimate of uncertainty in the jet-vetoed cross
section is the JVE method~\cite{Banfi:2013eda}.
The premise of the method is that the zero-jet cross section is given by
the product of the total cross section and jet-veto efficiency and
that the uncertainties in the two quantities are largely
uncorrelated. 
The argument that motivates this working assumption is that, at small
$p_{t}$, uncertainties in the efficiency are due to non-cancellation
of real and virtual contributions, while those in the total cross
section are connected with the large $K$-factor that is observed in
going from leading order to higher orders.

The JVE method can be applied both at fixed order and with
resummation.\footnote{In this respect it differs from the
  Stewart-Tackmann method~\cite{Stewart:2011cf}, which has so far been applied only to
  fixed-order calculations. An alternative way to estimate the
  theoretical uncertainties in the resummed case was proposed in~\cite{Stewart:2013faa}.}
It is useful to first extract the jet-veto efficiency from the total
Higgs cross section $\sigma_{\text{tot}}$ and the cross section
$\Sigma(p_{\rm t, veto})$ for Higgs production with a jet veto
(i.e.\ without any jets with $p_t > \ptjv$).
We define the expansion of the total
cross section and of the jet-veto cross-section up to
perturbative order ${\cal O}(\as^{2+n})$ as
\begin{equation}
  \label{eq:3}
  \sigmatot{n}= \sum_{i=0}^n \sigmai{i}\,,\qquad \Sigma(p_{\rm
    t,veto})=\sigmai{0}+\sum_{i=1}^n\Sigma^{(i)}(\ptjv)\,.
\end{equation}
Furthermore we use $ \bar{\Sigma}(\ptjv)$ to denote (minus) the cross
section to have at least one jet above a scale $\ptjv$.
Its order $\as^{2+i}$ component is  given by
\begin{align}
 \bar{\Sigma}^{(i)}(p_{\rm t, veto}) = -\int_{p_{\rm t, veto}}^{\infty}~dp_{\rm t}\frac{d\Sigma^{(i)}(p_{\rm t})}{dp_{\rm t}}.
\end{align}
This is related to $\Sigma^{(i)}(\ptjv)$ via 
\begin{align}
 \Sigma^{(i)}(p_{\rm t, veto}) = \sigmai{i} + \bar{\Sigma}^{(i)}(p_{\rm t, veto})\,.
\end{align}
From the above equations it is evident that one can obtain
$\Sigma^{(i)}(p_{\rm t, veto})$ at a given order in $\alpha_s$ by
combining the inclusive cross-section and the $H+1$ jet cross-section,
both computed at the same order in $\alpha_s$. Recently the $i=3$
coefficient was computed both for $\sigma_{\rm
  tot}$~\cite{Anastasiou:2015ema} and for 
$\bar{\Sigma}^{(i)}(p_{\rm t,
  veto})$~\cite{Boughezal:2015dra,Boughezal:2015aha,Caola:2015wna}.

The most obvious definition for the jet-veto efficiency
$\epsilon(\ptjv)$ is to write it as a ratio $\Sigma(\ptjv)/\sigma_{\rm tot}$ using
the highest order available in each case.
We call this prescription ``$(a)$'' and at N$^3$LO it reads
\begin{subequations}
\label{eq:epsilon_match_all}
\begin{equation}
  \label{eq:match_a}
  \epsilon^{(a)}(p_{\rm t, veto}) = 1+\frac{1}{\sigmatot{3}}\sum_{i=1}^3\bar \Sigma^{(i)}(p_{\rm t, veto})\,.
\end{equation}
In earlier work~\cite{Banfi:2012yh,Banfi:2012jm}, it had been argued
that in order to estimate perturbative uncertainties, one should
explore all possible ways of writing the series for $\epsilon(\ptjv)$
that retain the desired perturbative accuracy.
For example at N$^3$LO one can introduce scheme $(b)$, as 
\begin{equation}
 \label{eq:match_b}
  \epsilon^{(b)}(p_{\rm t, veto}) =1+\frac{1}{\sigmatot{2}}\sum_{i=1}^3\bar \Sigma^{(i)}(p_{\rm t, veto})\,,
\end{equation}
\end{subequations}
which is equivalent to scheme $(a)$ up to ${\cal O}(\alpha_s^4)$
corrections. Three further schemes are possible at N$^3$LO, where one
progressively expands $\sigma_{\rm tot}$ in the denominator while
ensuring the correctness of the full expression at N$^3$LO (see
Appendix~\ref{app:schemes}).
This in effect corresponds to using the degree of convergence for each
and every one of the previous orders as an input to determining the
possible size of unknown N$^4$LO corrections.
In the case at hand, however, the early terms of the series show
extremely poor convergence, especially at higher energies and when
including quark-mass effects.
As a result, taking an envelope of all possible schemes leads to
uncertainty estimates that grow very large, and contrast with the
good convergence observed in practice for the last known order
of both the numerator and the denominator.
A careful study of this question, summarised in Appendix~\ref{app:schemes}, has led us
to conclude that it is more appropriate to limit oneself to schemes
that give sensitivity to the convergence of just the last order of
the perturbative series.
This implies that we should take just the two schemes $(a)$ and
$(b)$ defined in~\eqref{eq:epsilon_match_all}.\footnote{When the
  prescription of Refs.~\cite{Banfi:2012yh,Banfi:2012jm} was
  originally introduced, only NNLO results were available.
  Their last order displayed still rather poor convergence, which
  justified a more conservative approach.}

Specifically, to estimate our uncertainty for a fixed-order
prediction, we will take the envelope of scheme $(a)$ with a 7-point
scale variation around a central scale $\mu_0$ ($\mu_{R,F}/\mu_0 =
\{\frac12,1,2\}$ with $\frac12 \le \mu_{R}/\mu_F \le 2$), together
with scheme $(b)$ evaluated at $\mu_{R,F} = \mu_0$.
The justification for not having scale variations in scheme $(b)$ is
that to include  them might effectively correspond to double
counting, i.e.\ summing two sources of uncertainty that may at some
level have shared origins.

In the results that follow (unless otherwise specified) we will
consider $13\TeV$ proton--proton collisions, with $R=0.4$ anti-$k_t$
jets~\cite{Cacciari:2008gp} as implemented in \textsc{FastJet}
v.~3.1.2~\cite{Cacciari:2011ma},
and use NNPDF2.3 parton distribution functions (PDFs) at
NNLO~\cite{Ball:2012cx}, accessed through
LHAPDF6~\cite{Buckley:2014ana}, with a strong coupling at the
$Z$-boson mass of $\as(M_Z)=0.118$. 
We choose $\mu_0 = m_H/2$ as the default renormalisation and
factorisation scale. No rapidity cuts are applied to the jets.

\begin{figure}
  \centering
  \includegraphics[width=0.49\textwidth]{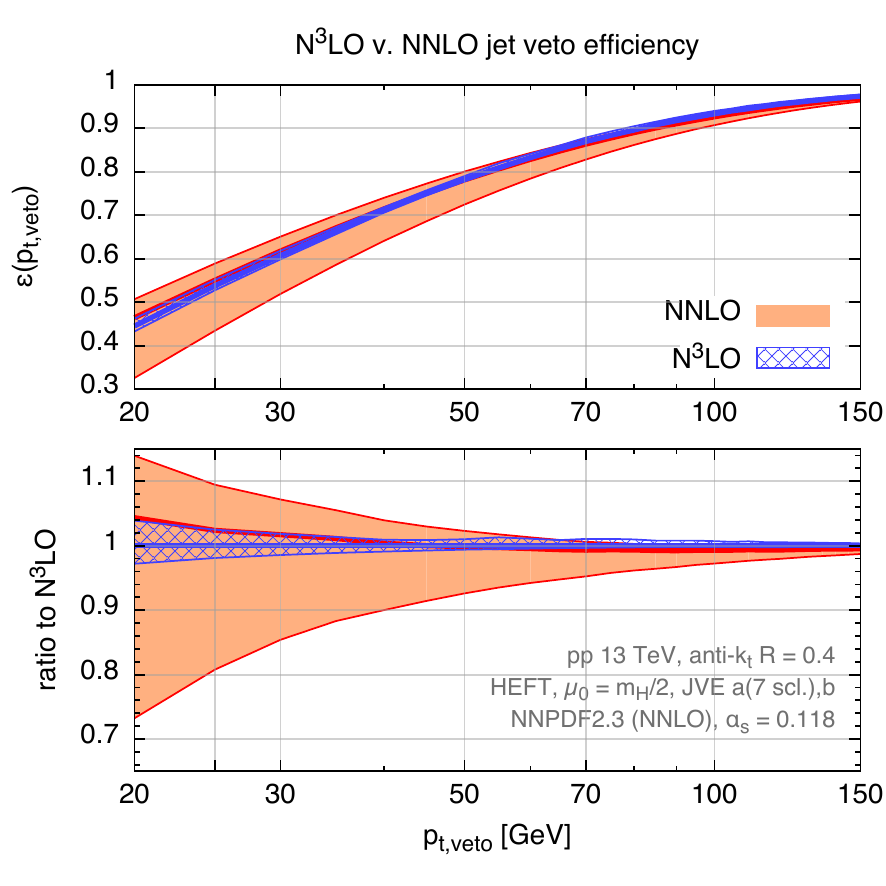}\hfill
  \includegraphics[width=0.49\textwidth]{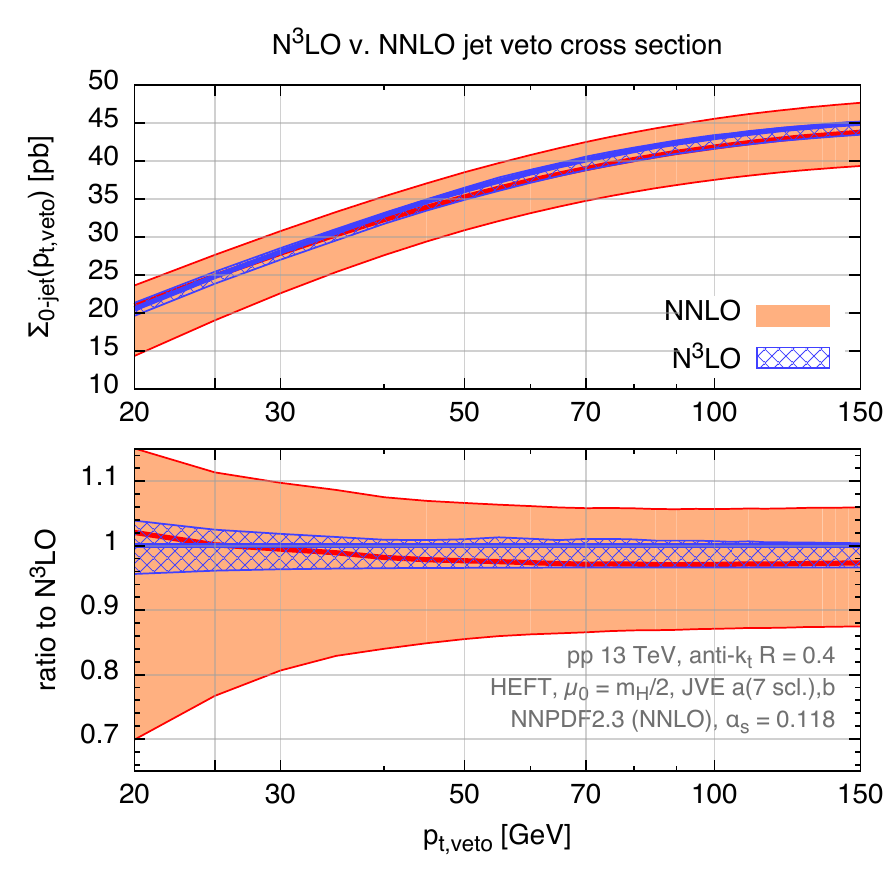}
  \caption{Comparison of NNLO and N$^3$LO results for the jet-veto
    efficiency (left) and the jet-veto cross section (right), using
    the updated jet-veto efficiency prescription described in
    section~\ref{sec:JVE}.
    The notation ``JVE a(7 scl.),b'' indicates the use of the jet
    efficiency methods with an uncertainty coming from the envelope
    of 7-point renormalisation and factorisation
    scale variation in scheme $(a)$ and additionally scheme $(b)$ with
    central scales.
  }
  \label{fig:n3lo-v-n2lo}
\end{figure}

A comparison of NNLO and N$^3$LO results with this prescription, in
the effective theory with a large top mass and no bottom mass, is shown in
Fig.~\ref{fig:n3lo-v-n2lo}, for both the jet-veto efficiency (left)
and the jet-veto cross section (right).
One observes a very considerable decrease in the uncertainties in going from
NNLO to \NNNLO with only a modest change in the central values,
associated with a small increase in the total cross section from the
\NNNLO corrections~\cite{Anastasiou:2015ema} and a slight decrease in
the low-$p_t$ efficiency associated with an increase in the 1-jet
cross section at NNLO~\cite{Boughezal:2015dra,Boughezal:2015aha,Caola:2015wna}.

\subsection{Resummation}

Next let us recall the structure of the NNLL resummed jet-veto cross
section~\cite{Banfi:2012jm},
\begin{multline}
  \label{eq:fullxsc}
  \Sigma_{\NNLL}(\ptjv)= \left( {\mathcal
      L}^{(0)}({L})+{\mathcal L}^{(1)}({L})\right) \times
  \\ \times
  \left (1+\mathcal{F}^{\text{clust}}(R)+\mathcal{F}^{\text{correl}}(R)\right)
  \times e^{ Lg_1(\as  L)+g_2(\as  L)+\frac{\as}{\pi} g_3(\as L)}\,,
\end{multline}
where we have split the factors involving the parton luminosities
into two terms ${\mathcal
  L}^{(0)}({L})$ and ${\mathcal L}^{(1)}({L})$, which
start at order $\alpha_s^2$ and $\alpha_s^3$ respectively:
\begin{align}
 \label{eq:L_0}
& {\mathcal L}^{(0)}({L}) = \sum_{i,j}\int dx_1 dx_2 |M_{B,ij}|^2\delta(x_1 x_2 s - M^2)
  f_i\!\left(x_1, e^{-{L}} \mu_F\right)f_j\!\left(x_2, e^{-{L}} \mu_F\right), \\ 
& {\mathcal L}^{(1)}({L}) =   \frac{\alpha_{s}}{2\pi}\sum_{i,j}\int
                           dx_1 dx_2 |M_{B,ij}|^2 \delta(x_1 x_2 s -
                           M^2) 
 \bigg[f_i\!\left(x_1, e^{-{L}} \mu_F\right)
  f_j\!\left(x_2, e^{-{L}} \mu_F\right){\cal H}^{(1)} +\notag\\
  &\frac{1}{1-2\alpha_s \beta_0 {L}}\sum_{k}\bigg(
  \int_{x_1}^1\frac{dz}{z} C_{ik}^{(1)}(z)
  f_k\!\left(\frac{x_1}{z}, e^{-{L}} \mu_F\right)
  f_j\!\left(x_2, e^{-{L}} \mu_F\right) +\{(x_1,i)\,\leftrightarrow\,(x_2,j)\}\bigg)\, \bigg]\,. 
\end{align}
Here $|M_{B,ij}^2|$ is the squared Born matrix element for the partonic
scattering channel $ij \to H$, $L \equiv \ln Q/\ptjv$ is the
logarithm we resum, where typically we choose the resummation scale
$Q$ of the order of $m_H/2$. ${\cal H}^{(1)}$ is a hard NLO correction,
$C^{(1)}_{ik}(z)$ is a NLO coefficient function and $f_i(x, \mu_F)$ is
the parton distribution function for flavour $i$ at factorisation
scale $\mu_F$.
The strong coupling $\as$ is always understood to be evaluated at a hard
scale $\mu_R \sim m_H/2$, 
$\beta_0 = (11C_A - 2n_f)/(12\pi)$, 
and the
factorisation scale $\mu_F$ is also to be chosen of the order of $m_H/2$.
The $g_i(\as L)$ functions encode the bulk of the LL, NLL and NNLL
resummation (for $i=1, 2, 3$ respectively). The $g_2$ and $g_3$
functions, as well as the ${\cal H}^{(1)}$ and $C^{(1)}$ coefficients all 
depend on the choice of $Q$. 
The quantities $\mathcal{F}^\text{clust}$ and
$\mathcal{F}^\text{correl}$~\cite{Banfi:2012yh} account for the NNLL
dependence of the result on the jet definition and are further
discussed below in section~\ref{sec:smallR}.
Explicit expressions for the above terms are to be found in the
supplementary material of Ref.~\cite{Banfi:2012jm}, and a number of the
elements are closely related to those derived for $p_t$
resummation~\cite{Bozzi:2005wk,Becher:2010tm}.

\subsection{Matching}
\label{sec:matching}

To put together the fixed-order and resummed results, we use matching
schemes that extend those presented in Ref.~\cite{Banfi:2012jm} to one
order higher. We refer the reader to that publication for a detailed 
explanation of the matching procedure. 
The matching schemes essentially correspond to the two schemes for the
fixed-order efficiency given in Eqs.~(\ref{eq:epsilon_match_all}).

To understand our prescriptions for matching, it is first instructive
to rewrite the fixed-order schemes for jet-veto efficiencies as ratios of
two cross sections:
\begin{equation}
  \label{eq:epsilon-as-ratio}
  \epsilon^{(x)}(\ptjv) \equiv
  \frac{\Sigma^{(x)}(\ptjv)}{\Sigma^{(x)}(\infty)}\,,
\end{equation}
where $\Sigma^{(x)}$ admits a different perturbative expansion for each
scheme $(x)$.
Specifically, each of the two fixed-order schemes of
Eq.~(\ref{eq:epsilon_match_all}) can be obtained by combining
Eq.~(\ref{eq:epsilon-as-ratio}) with one of the following
prescriptions for $\Sigma$:
\begin{subequations}
\begin{align}
  \label{eq:sigma_match_a}
  \Sigma^{(a)}(p_{\rm t, veto}) &=
  \sigmai{0}+ \Sigma^{(1)}+ \Sigma^{(2)}+\Sigma^{(3)}\,,
   \\
 \label{eq:sigma_match_b_fo}
  \Sigma^{(b)}(p_{\rm t, veto}) &= \sigmai{0}+ \Sigma^{(1)}+ \Sigma^{(2)}+\bar\Sigma^{(3)}\,.
\end{align}
\end{subequations}
Scheme $(a)$ is the exact expansion for $\Sigma$ and it trivially
gives $\epsilon^{(a)}(\ptjv)$.
For scheme $(b)$, observe that it is simply obtained by multiplying
$\epsilon^{(b)}$ by $\sigmatot{2}$.\footnote{
One could instead arrange for each of the $\Sigma^{(x)}$ to have
the property that it tends to $\sigmatot{3}$ for $\ptjv \to
\infty$, however this would complicate the expressions without bringing
any actual change in the final results for the jet veto efficiency and
cross section.}

A further standard element that we need is a modification of the
resummation so that its effect switches off for $\ptjv \gtrsim m_H$.
We do this by replacing $L \to \tilde L$, defined as
\begin{equation}
  \label{eq:Ltilde}
  \tilde L = \frac{1}{p} \ln \left(\left(\frac{Q}{\ptjv}\right)^p +
    1\right)\,. 
\end{equation}
The choice of $p$ is somewhat arbitrary and as in earlier
work~\cite{Banfi:2012yh} we take a fairly large value, $p=5$, to
reduce the residual contribution from resummation at high $p_t$.

For the matched cross-sections we obtain the following results: 
\begin{subequations}
\begin{align}
    \label{eq:matcha}
\Sigma^{(a)}_{\rm matched}&(p_{\rm t, veto}) =\frac{\Sigma_{\rm NNLL}(p_{\rm t, veto})}{\sigmai{0}(1+\delta\mathcal{L}(\tilde{L}))}
 \bigg[\sigmai{0}\left(1+\delta\mathcal{L}(\tilde L)\right)+\Sigma^{(1)}(p_{\rm t, veto})
 -\Sigma_{\rm NNLL}^{(1)}(p_{\rm t, veto})\nonumber\\
 &+\Sigma^{(2)}(p_{\rm t, veto})-\Sigma_{\rm NNLL}^{(2)}(p_{\rm t,
   veto})
+\Sigma^{(3)}(p_{\rm t, veto})-\Sigma_{\rm NNLL}^{(3)}(p_{\rm t, veto})
\nonumber \\&+\left(\delta{\mathcal L}^{(1)}(\tilde L)
 -\frac{\Sigma_{\rm NNLL}^{(1)}(p_{\rm t, veto})}{\sigmai{0}}+\delta{\mathcal
     L}^{(2)}(\tilde L)-\frac{\Sigma_{\rm NNLL}^{(2)}(p_{\rm t,
       veto})}{\sigmai{0}}\right)\nonumber \\& 
 \times \left(\Sigma^{(1)}(p_{\rm t, veto})-\Sigma_{\rm NNLL}^{(1)}(p_{\rm t,
     veto})\right)
+\left(\delta{\mathcal L}^{(1)}(\tilde L)-\frac{\Sigma_{\rm NNLL}^{(1)}(p_{\rm t, veto})}{\sigmai{0}}\right) \nonumber\\
&\times\left(\Sigma^{(2)}(p_{\rm t, veto})-\Sigma_{\rm NNLL}^{(2)}(p_{\rm t,
     veto})\right)
-\frac{\Sigma_{\rm NNLL}^{(1)}(p_{\rm t, veto})}{\sigmai{0}}\nonumber\\
&\times \left(\delta{\mathcal L}^{(1)}(\tilde L)-\frac{\Sigma_{\rm
      NNLL}^{(1)}(p_{\rm t, veto})}{\sigmai{0}}\right)  
 \left(\Sigma^{(1)}(p_{\rm t, veto})-\Sigma_{\rm NNLL}^{(1)}(p_{\rm t,
     veto})\right)
\bigg]\,,
  \\
    \label{eq:matchb}
  \Sigma^{(b)}_{\rm matched}&(p_{\rm t, veto}) =
     \Sigma^{(a)}_{\rm matched}(p_{\rm t, veto}) 
      - \sigmai{3} \frac{\Sigma_{\rm NNLL}(p_{\rm t,
                                                veto})}{\sigmai{0}(1+\delta\mathcal{L}(\tilde{L}))}\,.
\end{align}
\end{subequations}
In the above expressions, $\Sigma_{\rm NNLL}^{(n)}(\ptjv)$ denotes the
${\cal O}(\as^n)$ contribution to the NNLL resummed result. 
The resummed cross section and its expansion are defined
in terms of the modified logarithms $\tilde{L}$ as defined in
Eq.~\eqref{eq:Ltilde}.
We have also introduced
$\delta \mathcal{L}=\mathcal{L}^{(1)}/\mathcal{L}^{(0)}$. This
quantity admits a perturbative expansion in powers of $\alpha_s$,
starting at order $\alpha_s$. We denote this expansion as
$\delta \mathcal{L} = \delta \mathcal{L}^{(1)} + \delta
\mathcal{L}^{(2)}+\dots$.
Note that $\delta\mathcal{L}^{(1)}$ does not actually depend on
$\tilde L$.
Note also that, as for the fixed-order schemes, the normalisation at
$\ptjv\to \infty$ is different for each matching scheme, in particular
$\Sigma^{(x)}_{\rm matched}(\infty) = \sigmatot{i}$ with $i = 3, 2$ for
$x = a, b$.
Using Eq.~(\ref{eq:epsilon-as-ratio}), one always recovers the correct
normalisation $\epsilon(\ptjv) \to 1$ for $\ptjv\to\infty$.

Since the matching schemes above are multiplicative, for small
$\ptjv$, any finite remainder in the square brackets is multiplied by
a Sudakov form factor, ensuring that the cross section and efficiency
vanish in the limit $\ptjv\to 0$, since $\Sigma_{\NNLL}(\ptjv)$
vanishes in this limit.\footnote{Note that this behaviour of
  $\Sigma_{\NNLL}$ can be altered when including the small-$R$
  resummation. This only happens for rather small $R$ values, and it
  is therefore not present for phenomenologically relevant values of
  the jet radius.}

\begin{figure}
  \centering
  \includegraphics[width=0.49\textwidth]{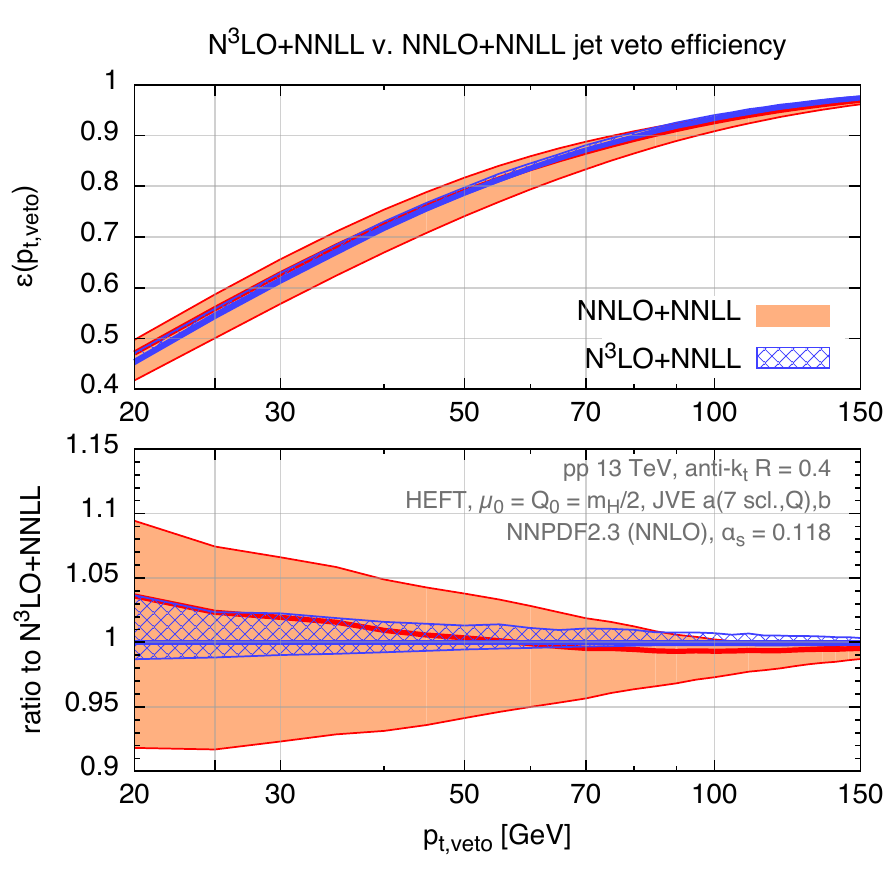}
  \hfill
  \includegraphics[width=0.49\textwidth]{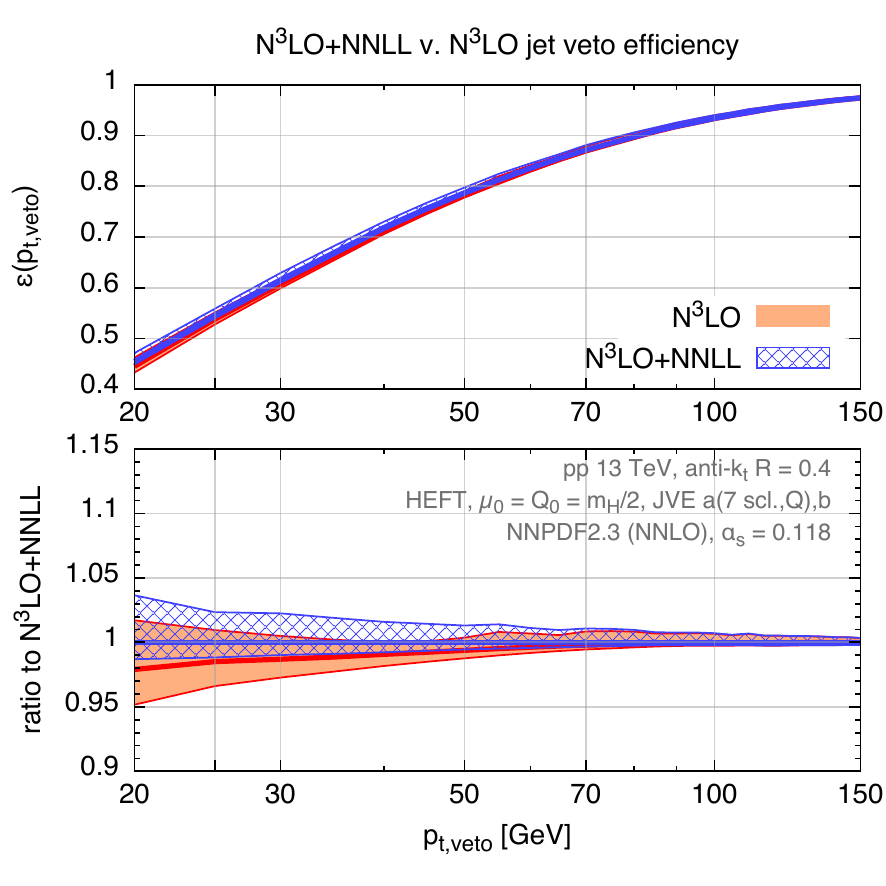}
  \caption{Comparison of matched N$^3$LO+NNLO results for the jet
      veto efficiency to NNLO+NNLL results (left) and to pure N$^3$LO
      predictions (right).}
  \label{fig:matching-impact}
\end{figure}

For matched results, in addition to varying $\mu_R$ and $\mu_F$ for
scheme $(a)$ and keeping a central choice for scheme $(b)$ (as done in
the fixed-order calculations), we also vary $Q$ in scheme $(a)$ around
its default choice of $Q_0 = m_H/2$.
However, in this work we change our convention for the range of $Q$
variation relative to earlier studies by some of
us~\cite{Banfi:2012yh,Banfi:2012jm}, which had $\frac12 \le Q/Q_0 \le 2$.
We instead choose the range of $\frac{2}{3}\le Q/Q_0 \le \frac32$
that had been originally proposed when $Q$ variation was first
introduced~\cite{Dasgupta:2002dc,Jones:2003yv}.
The motivation for returning to this earlier, narrower range comes
from the observation of the uncertainties at NNLO+NNLL: with the wider
range, the NNLO+NNLL uncertainties come out as unduly large relative
the actual changes observed when including N$^3$LO
corrections. Moreover, the old variation range gives rise to overly
large uncertainties in the tail of the leading jet's transverse
momentum differential spectrum. For a more detailed discussion of this
we refer the reader to Appendix~\ref{app:schemes}.

Fig.~\ref{fig:matching-impact} shows the impact of matching the NNLL
resummed results with the N$^3$LO result, compared to NNLO+NNLL
results (left) and to pure N$^3$LO results (right). 
In the left-hand plot, one sees a clear reduction in uncertainties in
going from NNLO+NNLL to N$^3$LO+NNLL, as expected given the impact of
the N$^3$LO results shown in Fig.~\ref{fig:n3lo-v-n2lo}.
While the NNLO+NNLL results had a substantially smaller uncertainty
band than pure NNLO, once one includes one additional order in $\as$,
resummation brings essentially no further reduction, as is visible in
the right-hand plot.
It does, however, induce a small shift in the central value (and
uncertainty band), whose magnitude is slightly smaller than the
uncertainty itself.

\subsection{Jet-radius dependence and small-$R$  effects}
\label{sec:smallR}

Two terms in Eq.~(\ref{eq:fullxsc}) are connected with the choice of
jet definition and in particular depend on the jet radius $R$.
$\mathcal{F}^\text{clust}(R)$ accounts for clustering of independent
soft emissions and for commonly used values of
$R$ is given by~\cite{Banfi:2012yh,Banfi:2012jm}
\begin{equation}
  \label{eq:fclust}
  \mathcal{F}^\text{clust}(R)=\frac{4\as^2(\ptjv) C_A^2 L}{\pi^2} \bigg(
  -\frac{\pi^2 R^2}{12} + \frac{R^4}{16}
  \bigg)\,.
\end{equation}
$\mathcal{F}^\text{correl}(R)$~\cite{Banfi:2012yh} comes from the
correlated part of the matrix element for the emission of two soft
partons.
For our purposes it is useful to further split it into two parts,
\begin{equation}
  \label{eq:fcorrel}
  \mathcal{F}^\text{correl}(R) =
  \frac{4\as^2(\ptjv)C_AL}{\pi^2}\left(
  f_1 \ln \frac{1}{R}
  +
  f_\text{reg}(R)\right)\,,
\end{equation}
where the coefficient of the logarithm of $R$ is

\begin{equation}
  \label{eq:fcorrel-smallR}
  f_1 = \frac{-131+12\pi^2+132\ln2}{72}C_A
                                   +\frac{23-24\ln2}{72}n_f\,,
\end{equation}
while the finite (regular) remainder is
\begin{equation}
  \label{eq:fcorrel-largeR}
  f_\text{reg}(R) \simeq 0.6106 C_A - 0.0155 n_f+\mathcal{O}(R^2)\,.
\end{equation}
This was originally derived including terms up to $R^6$ in
Ref.~\cite{Banfi:2012yh} with a numerically-determined constant term,
while an analytic form for the constant term and an expansion up to
order $R^{10}$ were given in Ref.~\cite{Becher:2013xia}.

Ref.~\cite{Tackmann:2012bt} advocated resummation of the terms
enhanced by powers of $\ln 1/R$. 
Ref.~\cite{Dasgupta:2014yra} showed that LL small-$R$ terms could be
incorporated into the jet-veto cross section by replacing
$\mathcal{F}^\text{correl}(R)$ with
\begin{multline}
  \label{eq:fcorrelmod}
  \mathcal{F}^\text{correl}_{\LLR}(R)=
  \exp \left[ - \frac{4\as(\ptjv) C_A}{\pi} L\, 
    {\cal Z}(t(R_0, R,\ptjv))
  \right]
  -1\\
  +\frac{4\as^2(\ptjv)C_A }{\pi^2} L \left(f_1 \ln \frac{1}{R_0} +
    f_\text{reg}(R) \right)\,, 
\end{multline}
where ${\cal Z}(t)$ (denoted
$\langle \ln z\rangle_g^\text{hardest}(t)$ in
Ref.~\cite{Dasgupta:2014yra}) is the \LLR resummed result for the
first logarithmic moment of the momentum fraction carried by the
hardest small-$R$ jet resulting from the fragmentation of a gluon.
A detailed, partially parametrised, expression for ${\cal Z}(t)$ is
given in Eq.~(\ref{eq:lnz}), with tabulated coefficients for
$n_f=4,\,5$ in table~\ref{tab:lnz-fit} (the second order coefficient
was also calculated in Ref.~\cite{Alioli:2013hba}).
The quantity $t(R_0, R,p_t)$ is an integral of the coupling over
scales related to the allowed emission angles, defined specifically as
\begin{equation}
  \label{eq:t-definition}
  t(R_0, R,\ptjv) =\int_{R^2}^{R_0^2} \frac{d\theta^2}{\theta^2}
  \frac{\as(\ptjv\theta)}{2\pi}\,.
\end{equation}
The nominally free parameter $R_0$ can be understood as a resummation
scale for the \LLR resummation, or, more physically, the largest
allowed emission angle. By default we will take $R_0 = 1$ and vary it
in the range $0.5 \le R_0 \le 2$.

In practice, in Eq.~(\ref{eq:fcorrelmod}) we will make the replacements
\begin{subequations}
  \begin{align}
    \as(\ptjv) &= \frac{\as}{1 - 2\lambda}\,,
    \\
    t(R_0, R,\ptjv) &= \frac{1}{2\pi \beta_0} \ln \frac{1-2\lambda}{1 - 2\lambda
                 - \as \beta_0 \ln \frac{R_0^2}{R^2}}\,,
                 \label{eq:t-homogeneous}
  \end{align}
\end{subequations}
where $\as \equiv \as(\mu_R)$, $\lambda \equiv \as \, \tilde L \, \beta_0$.
One sees explicitly from Eq.~(\ref{eq:t-homogeneous}) that logarithms
of $\ptjv$ (in $\lambda$) and $R$ are being treated on the same
footing, i.e.\ one is including all terms
$(\as \ln \ptjv)^m (\as \ln R)^n$ for any $m$ and $n$.
The expression includes just the logarithms needed to obtain joint
NNLL+\LLR resummation, without terms that are subleading in this
hierarchy (except for those explicitly included as part of a NNLL
resummation).\footnote{This ``minimal'' prescription is standard in
  resummations, however in the limit of sufficiently small $R$ and
  small $\ptjv$ we have observed certain artefacts that, as far as we
  understand, can only be cured by including subleading terms.
  Furthermore, inspecting the formulae, one immediately sees that
  the combination of small-$R$ and small-$\ptjv$ resummation may cause
  difficulties, since the smallest physical scale in the problem is
  now $R \ptjv$, which for sufficiently small $R$ can approach
  non-perturbative values even when $\ptjv \gg \Lambda_{QCD}$. For
  commonly-used values of the jet radius $R$ and $\ptjv$ the resummed cross
  section does not feature this issue, which is irrelevant for the
  phenomenology shown here.  Hence we leave the further study of this
  question to future work. We thank especially Mrinal Dasgupta for
  collaboration on this and related aspects.}

\begin{figure}
  \centering
  \includegraphics[width=0.49\textwidth]{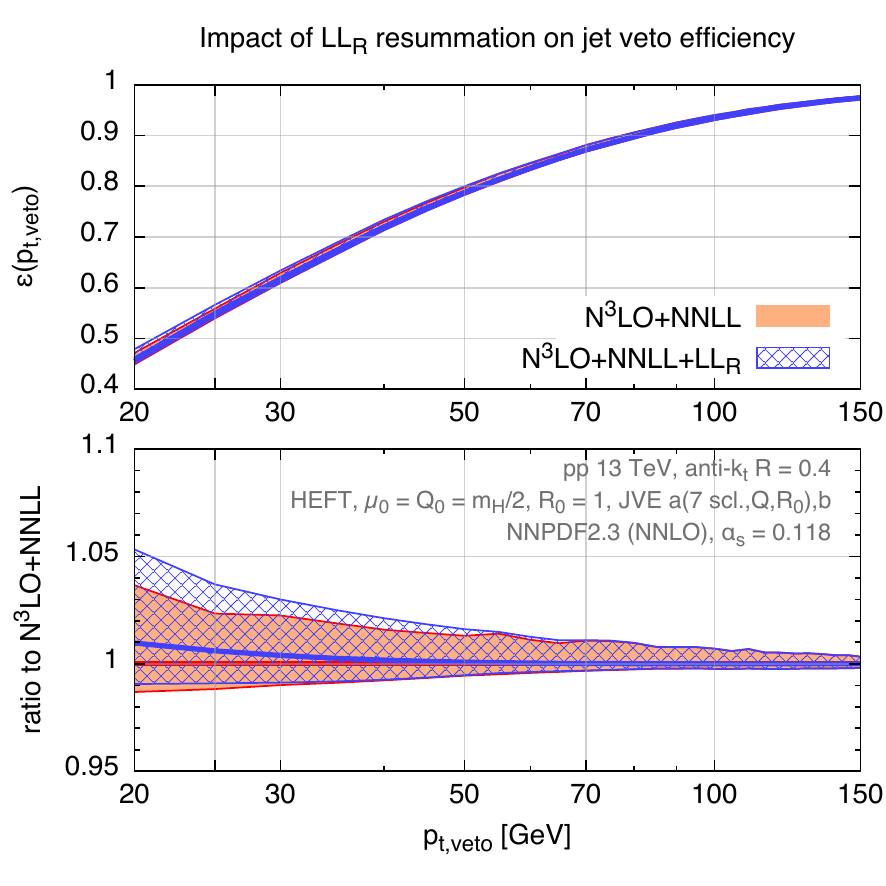}
  \caption{Impact of small-$R$ resummation on the jet-veto
    efficiency, comparing N$^3$LO+NNLL+\LLR to N$^3$LO+NNLL results.}
  \label{fig:lnR-impact}
\end{figure}

The impact of the small-$R$ resummation is shown in
Fig.~\ref{fig:lnR-impact}, where one sees that it increases the
central value of the efficiency by about $1\%$ at $\ptjv = 20\GeV$,
with a slight increase also in the size of the uncertainty band.
While it makes sense to include the \LLR resummation with a view to
providing the most complete prediction possible, for current
phenomenological choices of $R$ it does not bring a large
effect.\footnote{Ref.~\cite{Alioli:2013hba}, using a second order
  calculation of $\cal Z$, had also found small-$R$ effects that were
  small.}

\subsection{Quark-mass corrections}
\label{sec:quark-mass-corrections}

So far we have considered Higgs production in the heavy-top
approximation. In this section we study the corrections due to finite
top and bottom masses in the loop. Following the procedure of
Ref.~\cite{Banfi:2013eda}, the effect of heavy-quark masses at NNLL
amounts to simply replacing both the Born squared matrix element
$|M_{B,ij}|^2$ and the corresponding one-loop virtual correction
${\cal H}^{(1)}$ with the ones accounting for the correct quark-masses
dependence (cf. section 4.1 of ref.~\cite{Banfi:2013eda}).

We match the NNLL prediction so defined to the N$^3$LO fixed-order
cross section where we use the exact mass dependence up to NLO, while
keeping the heavy-top approximation for both NNLO and N$^3$LO
corrections. We use this as our default prescription for the results
presented below.
Moreover, we allow for different resummation scales for top and
bottom-induced effects. Therefore, we associate to bottom-induced
effects (mainly top-bottom interference) an additional resummation
scale $Q_b$. The matched cross section, including quark-mass effects
then reads
\begin{equation}
  \label{eq:Sigma-matched-q}
  \Sigma_{\rm matched}(\ptjv) =  \Sigma_{\rm matched}^{t}(\ptjv,Q)+\Sigma_{\rm matched}^{t,b}(\ptjv,Q_b)-\Sigma_{\rm matched}^{t}(\ptjv,Q_b)\,.
\end{equation}

\begin{figure}
  \centering
  \includegraphics[width=0.49\textwidth]{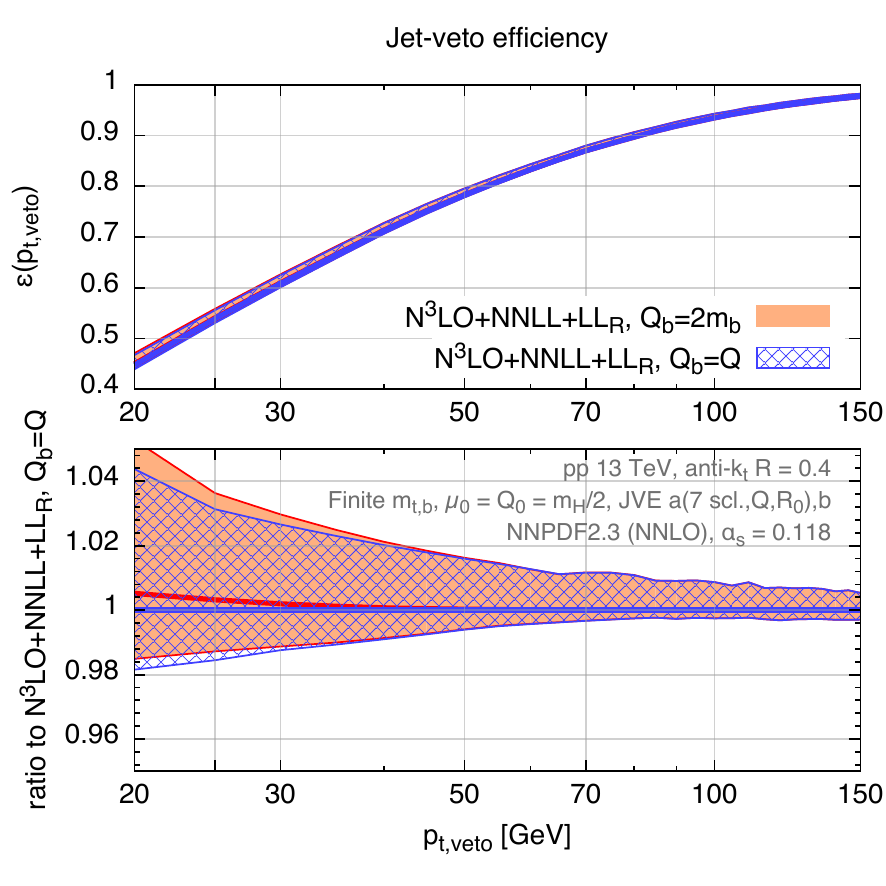}\hfill
  \includegraphics[width=0.49\textwidth]{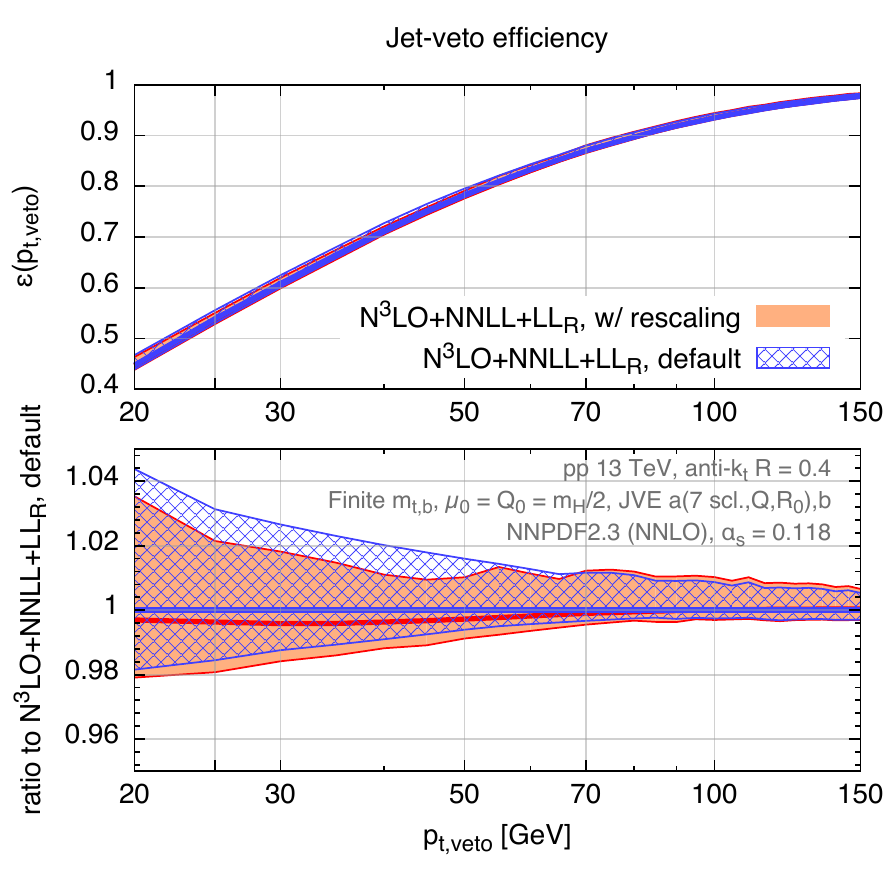}
  \caption{Left: the plot shows the impact of a different resummation
    scale for the bottom-induced contributions. Our default choice
    $Q_b=Q$ (in blue/hatched), and it is compared to the result with
    $Q_b=2\,m_b$ (in red/solid). See the text for description. Right:
    in the plot we compare two different ways of implementing finite
    quark-mass effects, as discussed in the text.}
  \label{fig:mass-effects}
\end{figure}

We set $Q$ to $Q_0=m_H/2$, and vary it as described in
section~\ref{sec:matching} ($\frac{2}{3} \le Q/Q_0 \le \frac{3}{2}$) to
estimate the associated uncertainty.  As far as $Q_b$ is concerned,
one could either set $Q_b=Q$ in the jet-veto efficiency (as done in
ref.~\cite{Banfi:2013eda}) or set it to small scales of the order of
$m_b$, as advocated in Ref.~\cite{Grazzini:2013mca}. 
As shown in~\cite{Banfi:2013eda}, if the resummation is matched
to (at least) NNLO, the impact of changing $Q_b$ is very moderate. We
show this feature in the left plot of Figure~\ref{fig:mass-effects}
where we compare the jet-veto efficiency obtained with a central
$Q_b=2 \,m_b$ to the one obtained with $Q_b=Q$. In order to be more
conservative in our test, we vary $Q_b$ by a factor of two in either
direction in the prediction obtained with $Q_b=2\,m_b$, while varying
it in the nominal range $\frac{2}{3} \le Q/Q_0 = Q_b/Q_0 \le \frac{3}{2}$
in the latter case.
To this order, the difference between the two prescriptions is
minimal.  We therefore decide to set $Q_b=Q$ as our central
resummation scales and vary them in a correlated way by a factor 3/2
up and down. With this choice the first and third term in the
r.h.s. of Eq.~\eqref{eq:Sigma-matched-q} cancel exactly.

In the context of top-mass corrections only, we note that one could
alternatively rescale the \NNLO{} and \NNNLO{} corrections by the
ratio of the Born cross section with exact top-mass dependence to the
corresponding heavy-top result. This rescaling is well justified in
the limit of emissions with a transverse momentum much smaller than
the top mass. In this region of the spectrum, the corrections to the
heavy-top approximation amount to a constant shift up to moderately
large $\ptjv$. This is indeed the region that contributes the most to
the total cross section, even more so when a jet veto is
applied. However, it is well known that this is not the case for
bottom-quark effects since the region where emissions are softer than
the bottom mass is strongly Sudakov-suppressed. This is reflected in
the non-trivial shape distortion of the spectrum at normal $\ptjv$
values~\cite{Bagnaschi:2011tu,Grazzini:2013mca,Banfi:2013eda}. 
Hence, in the small $\ptjv$ region the above rescaling does
not provide a reliable assessment of finite-mass effects.

While it is beyond the scope of this article to give a precise
assessment of higher-order mass effects, one can get a rough estimate
of their possible impact by comparing our default prescription to the
one where one rescales both NNLO and N$^3$LO corrections as discussed
above to include finite top-mass effects. We show this in the
right-hand plot of Figure~\ref{fig:mass-effects}. We observe very
moderate effects of the rescaling down to $\ptjv = 20$ GeV. This
statement is clearly not conclusive, and a more careful study is
necessary. Eventually, the issues related to quark-mass effects can
only be fixed once a NNLO calculation of mass-effects will be
available.

\section{N$^3$LO+NNLL+LL$_{R}$ cross section and 0-jet efficiency at 13 TeV}
\label{sec:all}

In this section we report our best predictions for the jet-veto
efficiency and cross section at the LHC. %
The various ingredients that we use were discussed in the previous
section, but for ease of reference we summarise them here:
\begin{itemize}
\item The total N$^3$LO cross section for Higgs production in gluon
  fusion~\cite{Anastasiou:2015ema}, obtained in the heavy-top
  limit.\footnote{The Wilson coefficient is expanded out consistently
    both in the computation of the total and the inclusive one jet
    cross section.}
\item The inclusive one-jet cross section at NNLO taken from the code
  of Ref.~\cite{Caola:2015wna}, in the heavy-top limit.
  In this computation the $qq$ channel is included only up to NLO, and
  missing NNLO effects are estimated to be well below scale variation
  uncertainties~\cite{Boughezal:2015aha}.
\item Exact top- and bottom-mass effects up to NLO in the
  jet-veto efficiency and cross section~\cite{Spira:1995rr}. Beyond
  NLO, we use the heavy-top result, as explained in
  section~\ref{sec:quark-mass-corrections}.
\item Large logarithms $\ln(Q/\ptjv)$ resummed to NNLL accuracy
  following the procedure of~\cite{Banfi:2012jm}, with the treatment of
  quark-mass effects as described in ref.~\cite{Banfi:2013eda}.
\item Logarithms of the jet radius resummed to LL accuracy,
  following the approach of ref.~\cite{Dasgupta:2014yra}. 
\end{itemize}

We consider 13 TeV LHC collisions with a Higgs-boson mass of
$m_H = 125$ GeV, compatible with the current experimental
measurement~\cite{Aad:2015zhl}. 
For the top and bottom pole quark masses, we use $m_t\,=\,172.5$ GeV and
$m_b\,=\,4.75$~GeV.
Jets are defined using the anti-$k_t$
algorithm~\cite{Cacciari:2008gp}, as implemented in \texttt{FastJet
  v3.1.2}~\cite{Cacciari:2011ma}, with radius parameter $R=0.4$, and
perform the momentum recombination in the standard $E$ scheme (i.e.\ summing the
four-momenta of the pseudo-particles).
We use NNPDF 2.3 parton distribution functions at NNLO with
$\alpha_s(m_Z) = 0.118$
(\texttt{NNPDF23\_nnlo\_as\_0118})~\cite{Ball:2012cx}.\footnote{Note
  that we use a five-flavour evolution for the running coupling as
  implemented in {\tt Hoppet}~\cite{Salam:2008qg}, while
  \texttt{NNPDF23\_nnlo\_as\_0118}~\cite{Ball:2012cx} uses a
  six-flavour scheme. This leads to small differences above the top
  threshold, which are however numerically irrelevant for our study.}  
In our central prediction for the jet-veto efficiency we
set renormalisation and factorisation scales to $\mu_R=\mu_F = m_H/2$.
The resummation scales are set to $Q= Q_b = m_H/2$,\footnote{$Q_b$
  applies when including top-bottom interference and bottom
  contributions, which we do by default here. As shown in
  section~\ref{sec:quark-mass-corrections}, switching to the
  alternative choice $Q_b = 2m_b$ makes less than $1\%$ difference.}
and we use matching scheme $(a)$~\eqref{eq:matcha} as default.
In this analysis, we do not include electro-weak
corrections~\cite{Actis:2008ug,Anastasiou:2008tj,Keung:2009bs}.

To determine the perturbative uncertainties for the jet-veto
efficiency we follow the procedure described in
section~\ref{sec:matching} and which we summarise here. 
We vary $\mu_R$,
$\mu_F$ by a factor of 2 in either direction, requiring $1/2 \le
\mu_R/\mu_F \le 2$.
Maintaining central $\mu_{R,F}$ values, we also vary $Q=Q_b$ in the
range $\frac{2}{3}\le Q/Q_0=Q_b/Q_0\le\frac{3}{2}$.
As far as the small-$R$ effects are concerned,
we choose the default value for initial radius for the evolution to
be $R_0=1.0$,\footnote{Note that it acts as a
  resummation scale for the resummation of logarithms of the jet
  radius. The initial radius for the small-$R$ evolution 
  differs from the jet radius used in the definition of jets, which is
  $R=0.4$.} and vary it conservatively by a factor of two in either direction. 
Finally, keeping all scales at their respective central
values, we replace the default matching scheme $(a)$~\eqref{eq:matcha}
with scheme $(b)$~\eqref{eq:matchb}. 
The final uncertainty band is
obtained as the envelope of all the above variations. 
We do not consider here the uncertainties associated with the parton
distributions (which mostly affect the cross section, but not the jet
veto efficiency), the value of the strong coupling or the impact of
finite quark masses on terms beyond NLO (which was discussed in
section~\ref{sec:quark-mass-corrections}).

We report the numerical values for our input total and one-jet cross section
in Table~\ref{tab:sigmatot}, with and without mass effects up to
${\cal O}(\alpha_s^3)$, with uncertainties obtained through scale
variation and using always NNLO PDFs and $\alpha_s$. 
\begin{table}
\begin{center}
{\renewcommand{\arraystretch}{1.1}
  \begin{tabular}{c||c|c||c|c||c|c}
LHC 13 TeV [pb]& $\sigmatot{2}$ &  $\sigmatot{3}$ & $\sigma_{\rm 1j \ge 25 {\rm GeV}}^{\rm NLO}$ &  $\sigma_{\rm 1j \ge 25 {\rm GeV}}^{\rm NNLO}$ & $\sigma_{\rm 1j \ge 30 {\rm GeV}}^{\rm NLO}$ &  $\sigma_{\rm 1j \ge 30 {\rm GeV}}^{\rm NNLO}$ 
    \\[0.2em] \hline 
EFT      & $45.1^{+4.0}_{-4.6}$ & $46.2^{+0.0}_{-1.6}$   & $20.3^{+3.6}_{-3.4}$ & $21.3^{+0.3}_{-1.3}$ & $17.3^{+3.0}_{-2.9}$ & $18.1^{+0.2}_{-1.1}$ \\
$t$-only & $47.1^{+4.3}_{-4.8}$ & $48.1^{+0.1}_{-1.9}$ & $20.7^{+3.8}_{-3.5}$ & $21.8^{+0.4}_{-1.4}$ & $17.6^{+3.2}_{-3.0}$ & $18.4^{+0.2}_{-1.2}$ \\
$t,b$    & $44.9^{+4.2}_{-4.7}$ & $45.9^{+0.0}_{-1.7}$   & $20.6^{+3.7}_{-3.5}$ & $21.6^{+0.4}_{-1.4}$ & $17.6^{+3.2}_{-3.0}$ & $18.4^{+0.2}_{-1.2}$
  \end{tabular}}
\end{center}
 \caption{Total cross section at NNLO ($\sigmatot{2}$) and
   at \NNNLO{} ($\sigmatot{3}$), and the one-jet
   cross-section $\sigma_{\rm 1j}$ at NLO and NNLO for central scales
   $\mu_0 = m_H/2$, with and without mass effects, as explained in the
   text. Uncertainties are obtained with a 7-point renormalisation and
   factorisation scale variation.
   Numbers determined from the computations of
   Refs.~\cite{Anastasiou:2015ema,Boughezal:2015dra,Caola:2015wna}. 
}
 \label{tab:sigmatot} 
\end{table}

\begin{figure}
  \centering
  \includegraphics[width=0.49\textwidth]{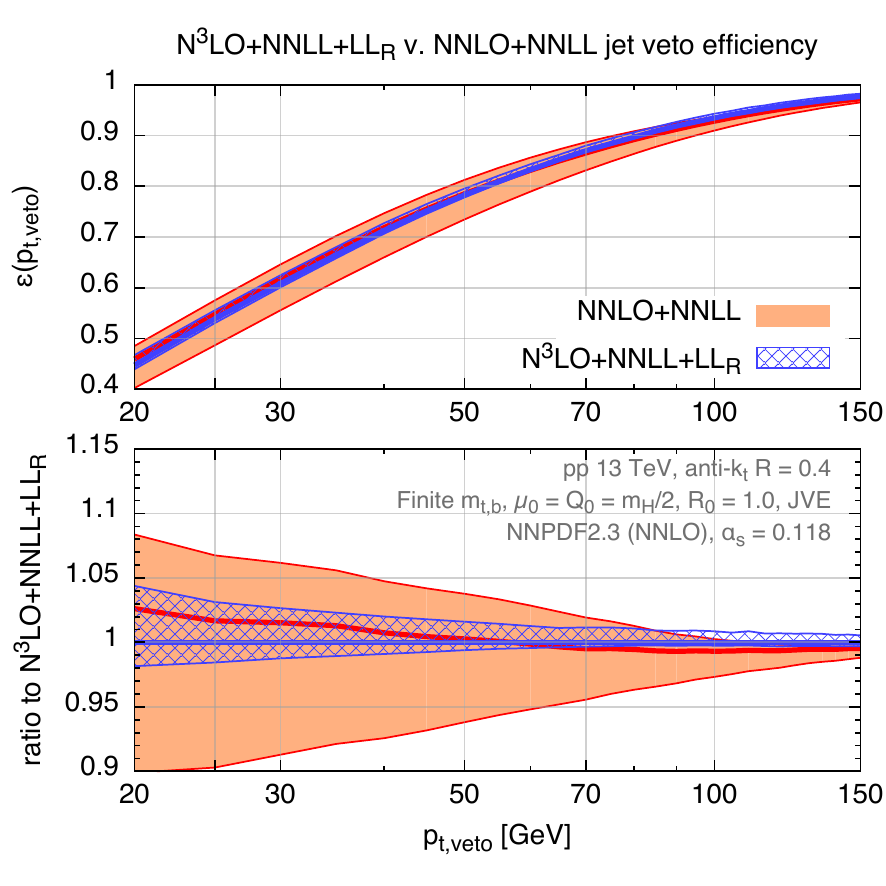}\hfill
  \includegraphics[width=0.49\textwidth]{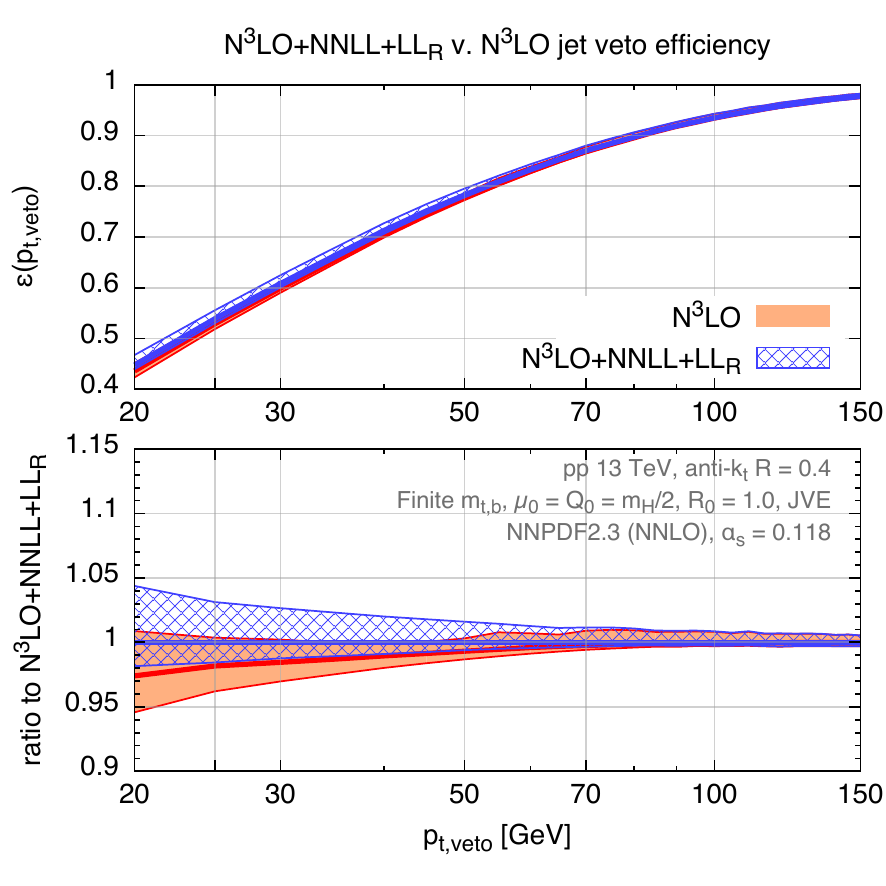}
  \caption{N$^3$LO+NNLL+LL$_R$ best prediction for the jet-veto efficiency (blue/hatched)
    compared to NNLO+NNLL (left) and fixed-order at N$^3$LO
    (right). 
  }
  \label{fig:bestprediction-efficiency}
\end{figure}
 Figure~\ref{fig:bestprediction-efficiency} (left) shows the
 comparison between our best prediction for the jet-veto efficiency
 (N$^3$LO+NNLL+LL$_R$) and  the previous NNLO+NNLL accurate prediction, both including mass effects. 
  We see that the impact of the N$^3$LO correction on the central value is of
 the order of 2\% at relevant jet-veto scales. The uncertainty band
 is significantly reduced when the N$^3$LO corrections are included,
 going from about 10\% at NNLO down to a few percent at N$^3$LO.
Figure~\ref{fig:bestprediction-efficiency} (right) shows the
comparison between the \NNLLNNNLO+\LLR prediction and the pure
N$^3$LO result. We observe a shift of the central value of the order
of 2\% for $\ptjv > 25\GeV$ when the resummation is turned on. 
In that same $\ptjv$ region, the uncertainty associated with the
N$^3$LO prediction is at the 2\% level, comparable with that of the
\NNLLNNNLO+\LLR prediction.
The fact that resummation effects are of
the same order as the uncertainties of the fixed order calculation
suggests that the latter might be accidentally small.
This situation is peculiar to our central renormalisation and
factorisation scale choice, $\mu_R = \mu_F = m_H/2$, and does not
occur at, for instance, $\mu_R=\mu_F=m_H$ (see
Appendix~\ref{app:mho2vsmh} for details).

\begin{figure}
  \centering
  \includegraphics[width=0.49\textwidth]{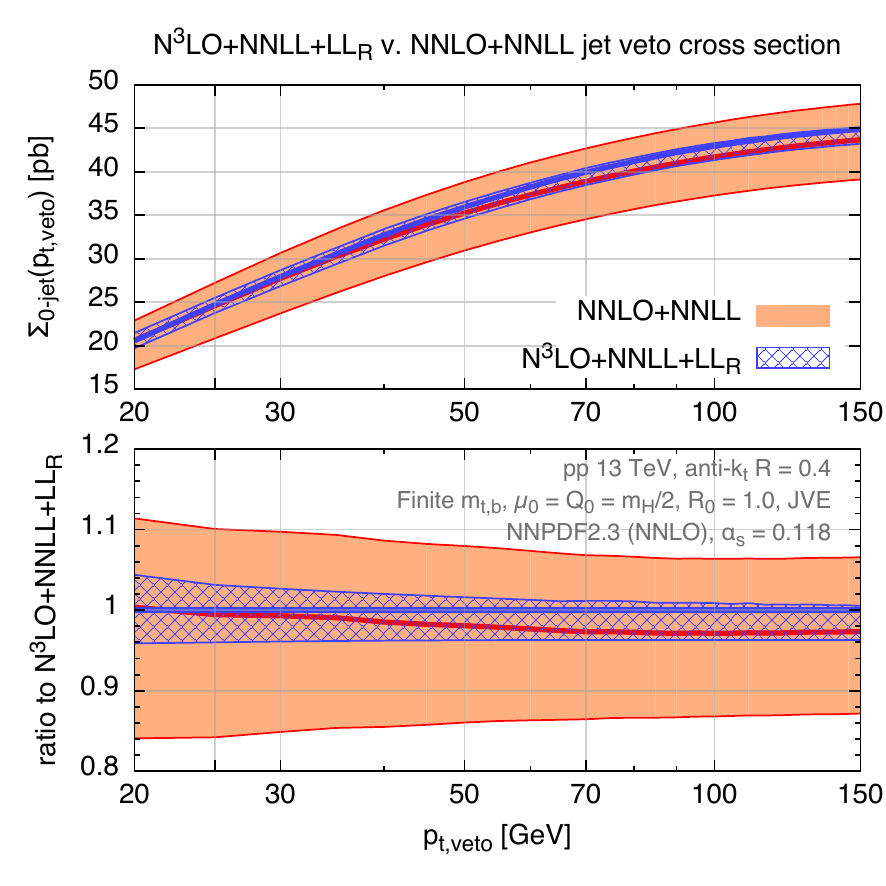}\hfill
  \includegraphics[width=0.49\textwidth]{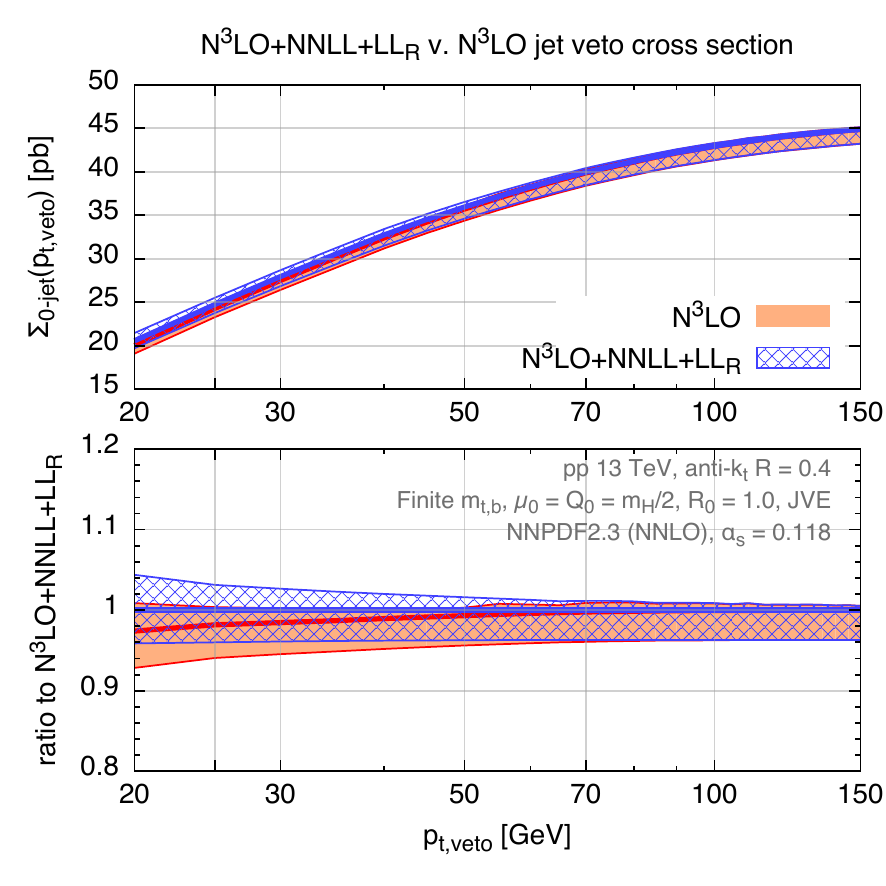}
  \caption{N$^3$LO+NNLL+LL$_R$ best prediction for the jet-veto cross section (blue/hatched)
    compared to NNLO+NNLL (left) and fixed-order at N$^3$LO
    (right).  }
  \label{fig:bestprediction-sigma}
\end{figure}

The zero-jet cross section is obtained as $\Sigma_\text{0-jet}(\ptjv)=\sigma_{\rm
  tot}\,\epsilon(\ptjv)$, and the inclusive one-jet cross section is
obtained as $\Sigma_{\ge\text{1-jet}}(p_{t,\rm{min}})=\sigma_{\rm
  tot}\,\left(1-\epsilon(p_{t,\rm{min}})\right)$.  The associated uncertainties
are obtained by combining in quadrature the uncertainty on the
efficiency obtained as explained above and that on the total cross
section, for which we use plain scales variations. The corresponding
results are shown in fig.~\ref{fig:bestprediction-sigma}. 
For this scale choice, we observe
that the effect of including higher-order corrections in the zero-jet
cross section is quite moderate at relevant $\ptjv$ scales. This is
because the small $K$ factor in the total cross section 
compensates for the suppression in the jet-veto efficiency. The
corresponding theoretical uncertainty is reduced by more than a factor
of two.

The predictions for jet-veto efficiency and the zero-jet cross section
are summarised in Table~\ref{tab:13-TeV-0jet}, for two experimentally
relevant $\ptjv$ choices. 
Results are reported both at fixed-order, and including the various
resummation effects.

\begin{table}
\begin{center}
{\renewcommand{\arraystretch}{1.1}
  \begin{tabular}{c|c|c|c|c}
    LHC 13 TeV
    & $\epsilon^{{\rm N^3LO+NNLL+LL_R}}$ & \,$\Sigma^{{\rm
                                N^3LO+NNLL+LL_R}}_\text{0-jet}\,{\rm [pb]}$\,  &
                                                          \,$\Sigma^{{\rm
                                                          N^3LO}}_\text{0-jet}$ &
                                                          \,$\Sigma^{{\rm
                                                          NNLO+NNLL}}_\text{0-jet}\,$
    \\[0.2em] \hline 
    $\ptjv=25\,{\rm GeV}$ & $0.539^{+0.017}_{-0.008}$ & $24.7^{+0.8}_{-1.0}$ & $24.3^{+0.5}_{-1.0}$ & $24.6^{+2.6}_{-3.8}$ \\
    $\ptjv=30\,{\rm GeV}$ & $0.608^{+0.016}_{-0.007}$ & $27.9^{+0.7}_{-1.1}$ &  $27.5^{+0.5}_{-1.1}$ & $27.7^{+2.9}_{-4.0}$        
  \end{tabular}}
\end{center}
 \caption{Predictions for the jet-veto efficiency and cross
   section at N$^3$LO+NNLL+LL$_{R}$, compared to the N$^3$LO and NNLO+NNLL cross
   sections.
   The uncertainty in the fixed-order prediction is obtained using the
   JVE method. All numbers include the effect of top and bottom quark
   masses, treated as described in the text, and are for a central scale $\mu_0=m_H/2$.
}
 \label{tab:13-TeV-0jet} 
\end{table}

\begin{figure}
  \centering
  \includegraphics[width=0.49\textwidth]{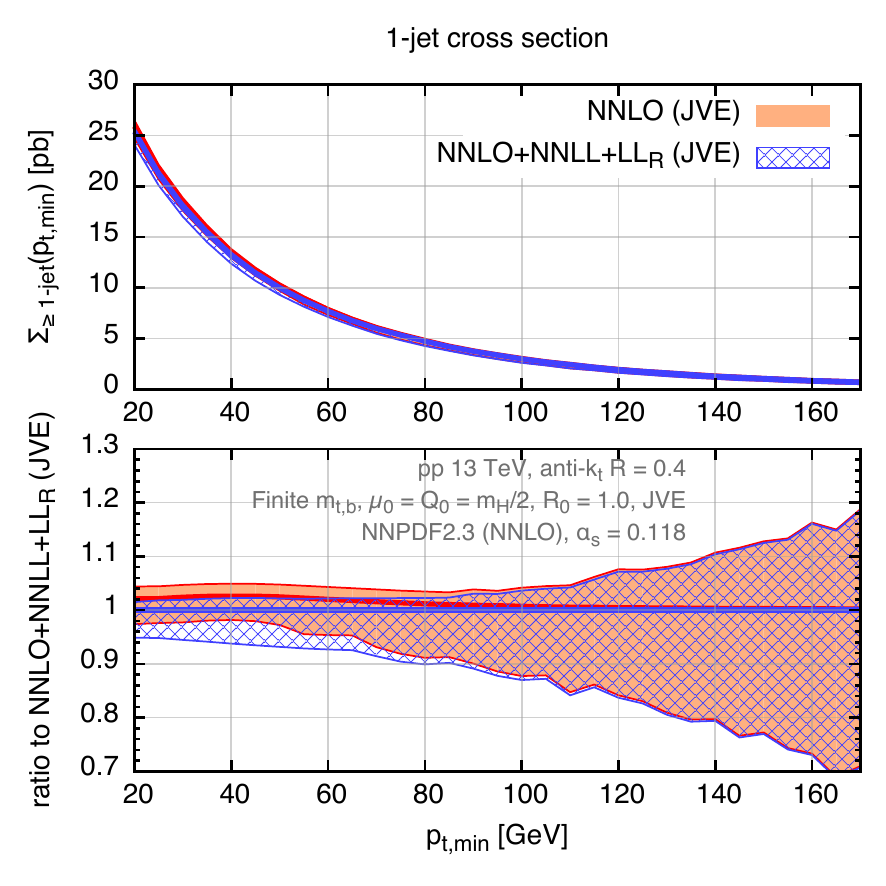}\hfill
  \includegraphics[width=0.49\textwidth]{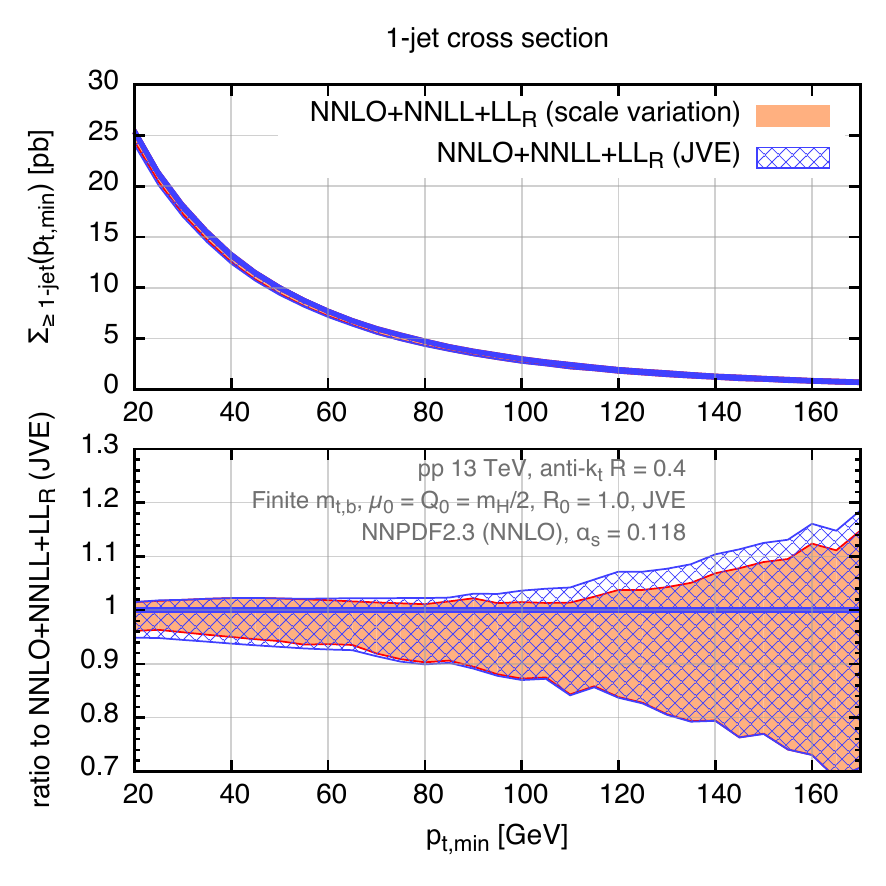}
  \caption{Matched NNLO+NNLL+\LLR prediction for the inclusive one-jet
    cross section (blue/hatched) compared to fixed-order at NNLO
    (left) and to the matched result with direct scale variation for
    the uncertainty (right), as explained in the text.}
  \label{fig:1jetxs}
\end{figure}

Figure~\ref{fig:1jetxs} shows the inclusive one-jet cross section
$\Sigma_{\ge \text{1-jet}}$, for which the state-of-the-art
fixed-order prediction is NNLO~\cite{Boughezal:2015dra,Boughezal:2015aha,Caola:2015wna}. The left-hand plot
shows the comparison between the best prediction at NNLO+NNLL+LL$_R$,
and the fixed-order at NNLO. Both uncertainty bands are obtained with
the JVE method outlined in Sec.~\ref{sec:matching}. We observe that the
effect of the resummation on the central value at moderately small
values of $\ptjv$ is at the percent level. Moreover, the inclusion
of the resummation leads to a slight increase of the theory
uncertainty in the small transverse momentum region.

The right-hand plot of Fig.~\ref{fig:1jetxs} shows our best prediction with
uncertainty obtained with the JVE method, compared to the case of just
scale (i.e.\ $\mu_R$, $\mu_F$, $Q$) variations. We observe a
comparable uncertainty both at small and at large transverse momentum,
which indicates that the JVE method is not overly conservative in the
tail of the distribution. We have observed that the same features
persist for the corresponding differential distribution.
Table~\ref{tab:13-TeV-1jet} contains the predictions for the
inclusive one-jet cross section for two characteristic $p_{t,\text{min}}$ choices.

\begin{table}
  \begin{center}
    {\renewcommand{\arraystretch}{1.1}
      \begin{tabular}{c|c|c}
        LHC 13 TeV
        &\, $\Sigma^{{\rm
          NNLO+NNLL+LL_R}}_{\ge \text{1-jet}}\,{\rm [pb]}$\,  &
                                                               \,$\Sigma^{{\rm
                                                               NNLO}}_{\ge\text{
                                                               1-jet}}\,{\rm
                                                               [pb]}$
        \\[0.2em]\hline
    $p_{\rm t, min}=25\,{\rm GeV}$ & $21.2^{+0.4}_{-1.1}$ & $21.6^{+0.5}_{-1.0}$\\
    $p_{\rm t,min}=30\,{\rm GeV}$ & $18.0^{+0.3}_{-1.0}$ & $18.4^{+0.4}_{-0.8}$
  \end{tabular}}
\end{center}
 \caption{Predictions for the inclusive one-jet cross section at
   NNLO+NNLL+LL$_{R}$ and NNLO. The uncertainty in the fixed-order
   prediction is obtained using the JVE method. All numbers include
   the effect of top and bottom quark masses, treated as described in
   the text, and are for a central scale $\mu_0=m_H/2$. }
 \label{tab:13-TeV-1jet} 
\end{table}

\section{Conclusions}
\label{sec:conclu}

In this article we have presented new state-of-the-art,
N$^3$LO+NNLL+LL$_R$, predictions for the jet-veto efficiency and the
zero-jet cross section in gluon-fusion induced Higgs production, as
well as NNLO+NNLL+LL$_R$ results for the inclusive one-jet cross
section.
The results, shown for 13\,TeV LHC collisions, incorporate recent
advances in the fixed-order calculation of the total cross
section~\cite{Anastasiou:2015ema}, the fixed-order calculation of the
one-jet cross
section~\cite{Boughezal:2015dra,Boughezal:2015aha,Caola:2015wna} and
the resummation of small-$R$ effects~\cite{Dasgupta:2014yra}.
They also include the earlier NNLL jet $p_t$
resummation~\cite{Banfi:2012jm} including finite quark mass
effects~\cite{Banfi:2013eda}.
Uncertainties have been determined using the jet-veto efficiency
method, which has been updated here to take into account the good
perturbative convergence observed with the new fixed-order
calculations.

Results for the jet-veto efficiency and zero-jet cross section for
central scale choices of $\mu_0=m_H/2$ and $\mu_0=m_H$ are reported in
tables~\ref{tab:13-TeV-0jet} and~\ref{tab:13-TeV-0jet-mh},
respectively. 
With our central scale choice, $\mu_0 = m_H/2$, we find that the
inclusion of the new calculations decreases the jet-veto efficiency by
2\% with respect to the NNLO+NNLL prediction, and it has a
substantially smaller uncertainty, reduced from more than 10\%
to less than 5\%. 

In the zero-jet cross section, the reduction in the
jet-veto efficiency is
compensated by a similar increase in the total cross section due to
the \NNNLO{} correction, resulting in a sub-percent effect. In
comparison to the \NNNLO{} result, the matched N$^3$LO+NNLL+LL$_R$
jet-veto efficiency and zero-jet cross section are about 2\% larger,
and have comparable ($\sim 3-4\%$) theoretical errors.
The picture is different for a central scale $\mu_0=m_H$, as
discussed in appendix~\ref{app:mho2vsmh}. In this case the jet-veto
efficiency at N$^3$LO+NNLL+LL$_R$ decreases by more than 5\% with
respect to the NNLO+NNLL result, while it is in perfect agreement with
the pure \NNNLO{} prediction. Perturbative uncertainties are
considerably (moderately) reduced with respect to the NNLO+NNLL
(\NNNLO{}) prediction.
For the inclusive one-jet cross section, we find a similar impact of
the resummation in the small $\ptjv$ region, and agreement with the
fixed-order scale variation at large transverse momentum values.

We stress that other corrections are of the same order as the
theoretical uncertainties obtained here. These involve electro-weak
effects, exact quark-mass treatment beyond the orders currently known,
and non-perturbative effects.
Furthermore, we stress that the results quoted here do not
account for PDF and strong coupling uncertainties, which also are at
the few-percent level.

Code for performing the resummation and matching with fixed order
predictions is publicly available in version~3 of the {\tt JetVHeto}
program~\cite{JetVHeto}.

\section*{Acknowledgements}

FC would like to thank Kirill Melnikov and Markus Schulze for
collaboration and exchanges concerning the Higgs plus one jet
computation. GPS and FAD would like to thank Mrinal Dasgupta, Matteo
Cacciari and Gregory Soyez for collaboration in the early stages of
the small-$R$ resummation part of this work. 
We wish to thank Babis Anastasiou, Claude Duhr and Bernhard
Mistlberger for interesting discussions regarding the total cross
section, and Kirill Melnikov for comments on the manuscript.
GZ is supported by the HICCUP ERC Consolidator grant.
PM is supported by the Swiss National Science Foundation (SNF) under
grant PBZHP2-147297.
The work of AB is supported by Science and Technology Facility Council
(STFC) under grant number ST/L000504/1.
FAD is supported by the ILP LABEX (ANR-10-LABX-63) financed by French
state funds managed by the ANR within the Investissements d'Avenir
programme under reference ANR-11-IDEX-0004-02.
GPS is supported in part by ERC Advanced Grant Higgs@LHC.
The work of FD is supported by the Swiss National Science Foundation
(SNF) under contract 200021-143781 and by the European Commission
through the ERC grant ``IterQCD''.  AB and PM wish to thank CERN for
hospitality, and FC, PM, GS and GZ would like to express a special
thanks to the Mainz Institute for Theoretical Physics (MITP) for its
hospitality and support while part of this work was carried out.

\appendix

\section{Revisited JVE uncertainty prescription}
\label{app:schemes}

In this paper we argued that the JVE method of
ref.~\cite{Banfi:2012yh,Banfi:2012jm} used to estimate uncertainties
should be modified. In this appendix we wish to motivate why we
revisited the JVE prescription.
We will argue that, while the original JVE method was appropriate
when it was proposed (i.e.\ when only the NNLO correction to the 0-jet
cross section in Higgs production was known), now that the \NNNLO{}
correction is available it would give rise to excessively conservative
uncertainties.

It is useful to first recall the original JVE method. 
In refs.~\cite{Banfi:2012yh,Banfi:2012jm}, to determine
uncertainties in the NNLO+NNLL prediction, $\mu_R$ and $\mu_F$ were
varied by a factor of 2 in either direction, requiring $1/2 \le
\mu_R/\mu_F \le 2$.  Maintaining central $\mu_{R,F}$ values, $Q$ was
also varied by a factor of 2 and changed the matching scheme, from the
scheme $(a)$ to schemes $(b)$ and $(c)$ as defined
in~\cite{Banfi:2012jm}.
The final uncertainty band was the envelope of these variations
(cf.~\cite{Banfi:2012yh}).
Our new prescription differs from the old one in two important points:
\begin{itemize}
\item only schemes $(a)$ and $(b)$ are used to probe the sensitivity to the matching scheme;
\item the range for the resummation scale variation is
  $2/3\le Q/Q_0\le 3/2$, as suggested originally in
  ref.~\cite{Dasgupta:2002dc}.
\end{itemize}
In the rest of this appendix we comment on both of these aspects.

The reason for having different schemes is that the efficiency is a
ratio of the jet-vetoed cross section to the total cross section.
Even at fixed order there is some freedom as to which
perturbative terms one chooses to keep in the denominator, or
alternatively expand out.
Different matching formulae can then be constructed that reproduce the
corresponding fixed-order expansions for the JVE efficiency.
At N$^3$LO, in addition to schemes $(a)$ and $(b)$ defined in
Eqs.~(\ref{eq:epsilon_match_all}), one can introduce three further schemes:
\begin{subequations}
\label{eq:epsilon_match_all-app}
\begin{align}
 \label{eq:match_c}
  \epsilon^{(c)}(p_{\rm t, veto}) &= 1+\frac{1}{\sigmatot{1}}\left[\sum_{i=1}^3\bar{\Sigma}^{(i)}(p_{\rm t, veto})
  -\frac{\sigmai{2}}{\sigmatot{0}}\bar
  \Sigma^{(1)}(p_{\rm t, veto})\right]\,,
\\
 \label{eq:match_cp}
  \epsilon^{(c')}(p_{\rm t, veto}) &= 1+\frac{1}{\sigmatot{1}}\left[\sum_{i=1}^3\bar{\Sigma}^{(i)}(p_{\rm t, veto})
  -\frac{\sigmai{2}}{\sigmatot{1}}\bar
  \Sigma^{(1)}(p_{\rm t, veto})\right]\,,
\\
 \label{eq:match_d}
  \epsilon^{(d)}(p_{\rm t, veto}) &=   1+\frac{1}{\sigmatot{0}}\left[\sum_{i=1}^3\bar{\Sigma}^{(i)}(p_{\rm t, veto})
  -\frac{\sigmai{1}}{\sigmatot{0}}(\bar{\Sigma}^{(1)}(p_{\rm t, veto})+\bar{\Sigma}^{(2)}(p_{\rm t, veto}))
 \nonumber  \right. \\& 
      \qquad\qquad\qquad \qquad\qquad\qquad
      \left.+\frac{\sigmai{1}\sigmai{1}-\sigmai{0} \sigmai{2}}{(\sigmatot{0})^2}\bar{\Sigma}^{(1)}(p_{\rm t, veto})\right]\,.
\end{align}
\end{subequations}
The schemes differ only by terms beyond \NNNLO.\footnote{Corresponding formulae for the matching
schemes can be found in the documentation of \tt{JetVHeto-v3}~\cite{JetVHeto}.}
At \NNLO{} there are just three schemes, $(a)$, $(b)$ and $(c)$, which
respectively have $\sigmatot{2}$, $\sigmatot{1}$ and $\sigmatot{0}$ in
the denominator:
\begin{subequations}
\label{eq:nnlo_epsilon_match_all}
\begin{align}
  \label{eq:nnlo_match_a}
  \epsilon^{(a)}_\NNLO(p_{\rm t, veto}) &= 1+\frac{1}{\sigmatot{2}}\sum_{i=1}^2\bar \Sigma^{(i)}(p_{\rm t, veto})\,,
  \\
  \label{eq:nnlo_match_b}
  \epsilon^{(b)}_\NNLO(p_{\rm t, veto}) &= 1+\frac{1}{\sigmatot{1}}\sum_{i=1}^2\bar \Sigma^{(i)}(p_{\rm t, veto})\,,
  \\
  \label{eq:nnlo_match_c}
  \epsilon^{(c)}_\NNLO(p_{\rm t, veto}) &=
  1+\frac{1}{\sigmatot{0}}\left[
    \sum_{i=1}^2\bar \Sigma^{(i)}(p_{\rm t, veto})
  -\frac{\sigmai{1}}{\sigmatot{0}}\bar{\Sigma}^{(1)}(p_{\rm t, veto})
  \right] ,
\end{align}
\end{subequations}
where, to avoid confusion, here we have explicitly added a ``NNLO''
label. 
In what follows, we will drop this label.\footnote{Note that there is
  a natural correspondence between \NNNLO and \NNLO schemes $(a)$
  and $(b)$.}

\begin{figure}
  \centering
  \includegraphics[width=0.49\textwidth]{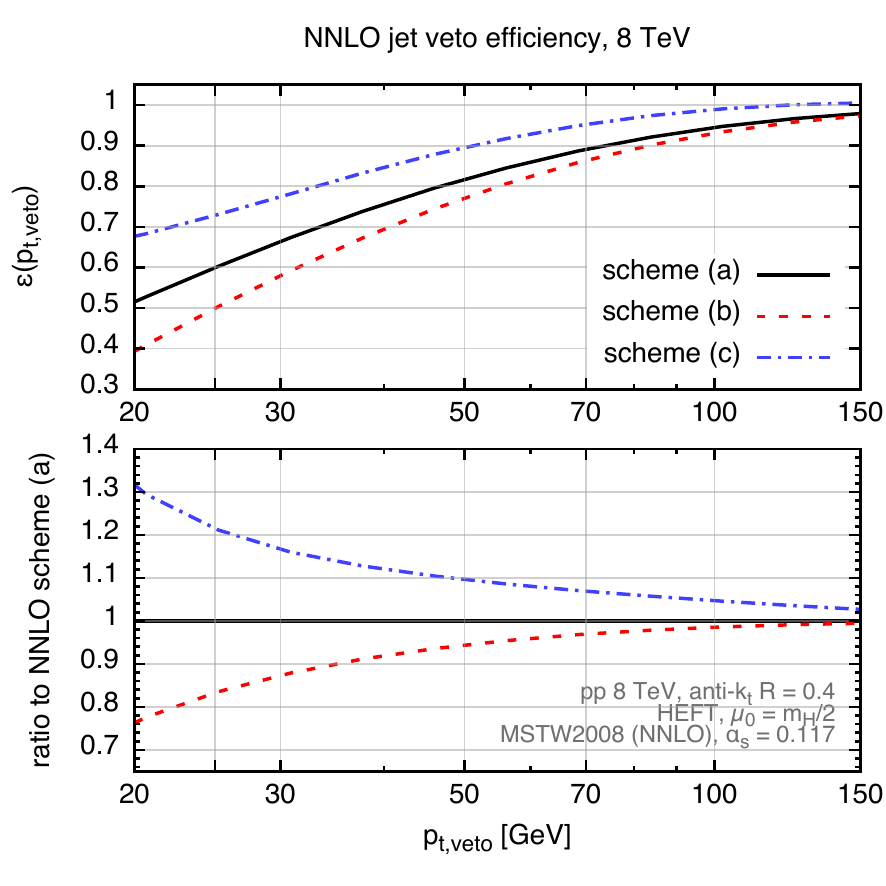}\hfill
  \includegraphics[width=0.49\textwidth]{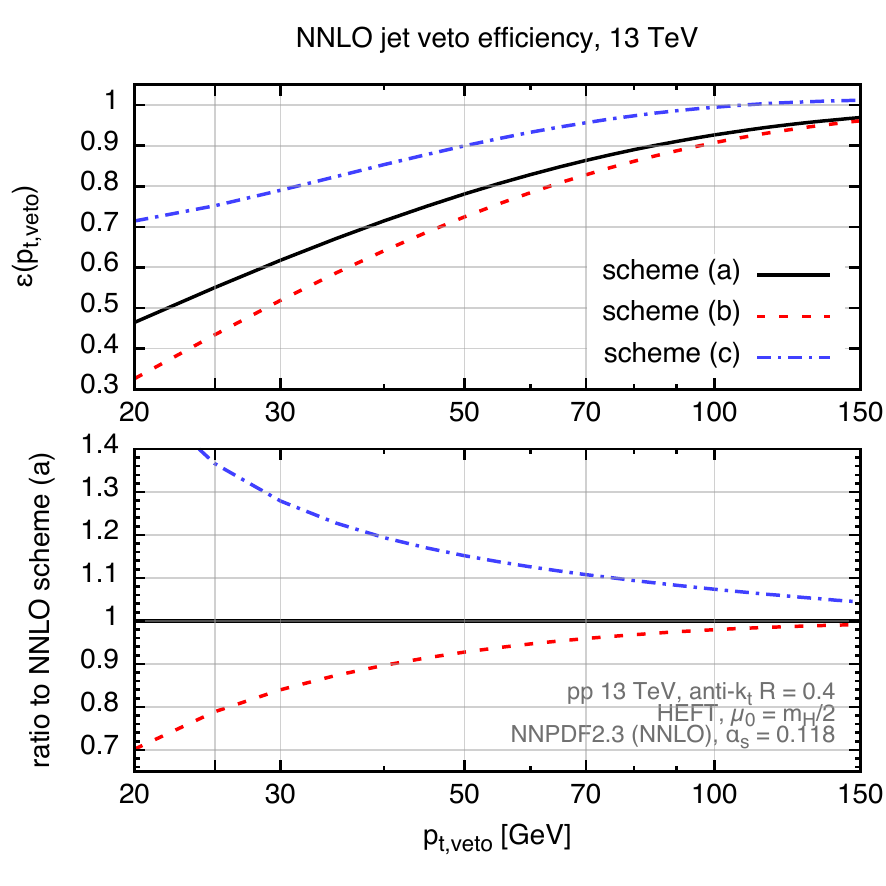}
  \caption{Comparison of the NNLO prediction for the jet-veto
    efficiency at 8\,TeV (left) and 13\,TeV (right). 
    The plots show the three efficiency schemes contributing to the
    uncertainty band in the old formulation of the JVE method.}
  \label{fig:oldJVE8vs13}
\end{figure}

To understand why we now restrict the scheme variation to schemes
$(a)$ and $(b)$, we first show in Fig.~\ref{fig:oldJVE8vs13} a
comparison between the NNLO jet-veto efficiency at 8\,TeV and 13\,TeV,
where we plot the three different possible matching schemes at this
order (for the central scale choice). 
Concentrating first on the absolute values of the efficiency, one sees
that in schemes $(a)$ and $(b)$ there is a reduction in going
from 8 to 13 TeV.
This is consistent with the expectation of an increase in the fraction
of events containing a jet when one goes to higher centre-of-mass
energy.
In contrast, the efficiency increases in scheme $(c)$.
This seems unphysical.
The combination of the different behaviours of schemes $(a)$ and $(c)$
has the consequence of a very substantial increase in apparent uncertainty.
Moreover, at sufficiently high $\ptjv$ scheme $(c)$ returns an
unphysical efficiency $\epsilon^{(c)}>1$.

The issues with scheme $(c)$ are to some extent understood, since
scheme $(c)$ at NNLO is very sensitive to the convergence of the first
correction in the perturbative expansion.
It is well-known that the first terms for the Higgs cross section
converge very poorly. 
In particular, the ratio of NLO to LO cross section contributions,
$\sigmai{1}/\sigmai{0}$, goes from about $1.23$ to $1.30$ between $8$
and $13\TeV$.\footnote{The $\sigmai{1}/\sigmai{0}$ ratio is further
  enhanced when mass effects are included.}
The difference between schemes $(a)$ and $(b)$, on the other hand, is
only sensitive to the size of the last perturbative order, hence we
believe it provides useful, but not overly conservative, information
on the uncertainty.

\begin{figure}
  \centering
  \includegraphics[width=0.49\textwidth]{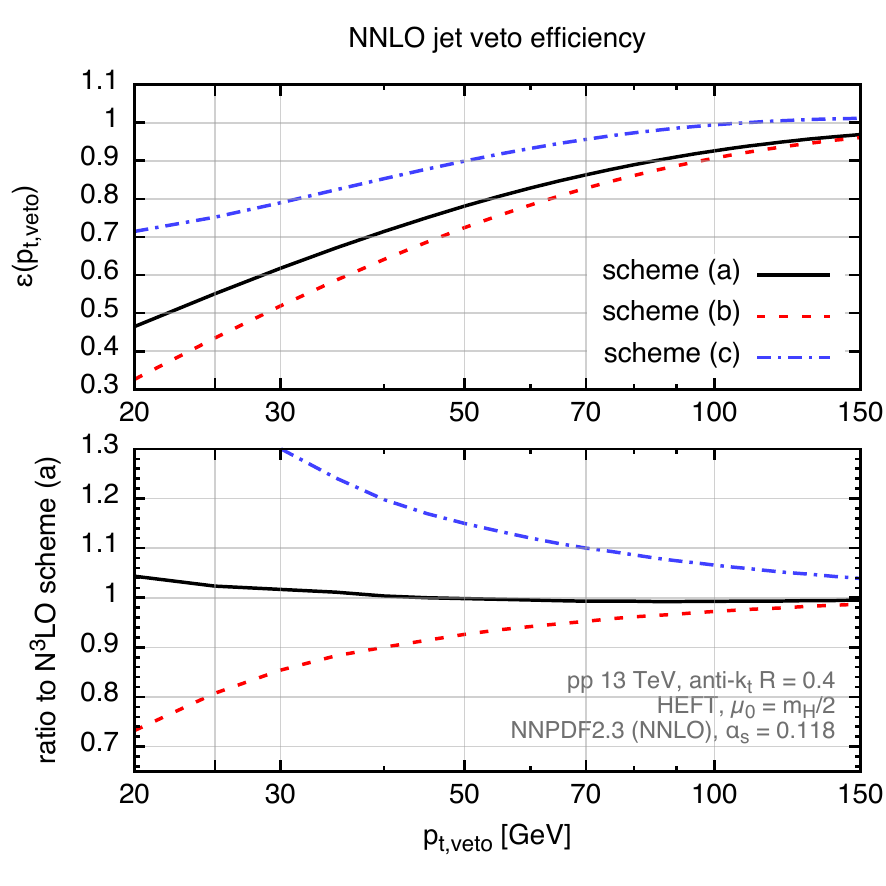}\hfill
  \includegraphics[width=0.49\textwidth]{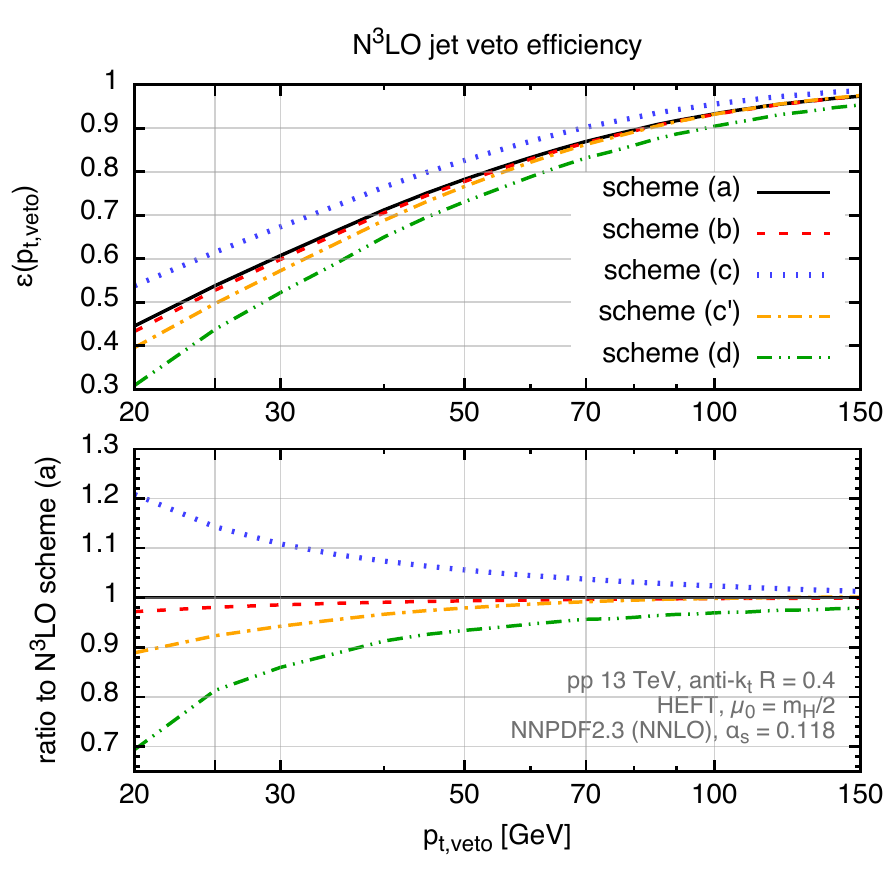}
  \caption{Jet-veto efficiency at 13\,TeV. The plots show the various
    efficiency schemes at \NNLO{} (left) and \NNNLO{} (right). The
    lower panels show the ratio to the \NNNLO{} central prediction
    (scheme $(a)$).}
  \label{fig:oldvsnewJVE}
\end{figure}

In order to study the impact of the new prescription for the
efficiency scheme variation, in Fig.~\ref{fig:oldvsnewJVE} we show the
fixed-order efficiency at \NNLO{} and \NNNLO{}, concentrating on the
13\,TeV case, where the
impact of scheme $(c)$ (at \NNLO{}) and $(d)$ (at \NNNLO{}) is more
pronounced. 
Fig.~\ref{fig:oldvsnewJVE} shows the various efficiency schemes
contributing at a given order according to the old JVE
prescription. 
We see that at \NNNLO{} the spread between schemes $(a)$ and $(b)$ is
comparable with the change in the efficiency from \NNLO{} to \NNNLO{},
while the inclusion of additional schemes $(c)$, $(c')$ and $(d$)
gives rise to a much larger uncertainty.
This suggests that the old JVE prescription is overestimating
uncertainties at this c.o.m. energy. 
This is even more true when including finite quark-mass effects (not
shown here).

 It is however also clear that, since the $(b)$ scheme prediction is
 obtained by computing the jet-veto efficiency at the central scale
 only, if the \NNNLO{} correction to the total cross section is
 accidentally very small at that scale, schemes $(a)$ and $(b)$ will
 return nearly identical values. Therefore, the corresponding scheme
 uncertainty will be very small. For our central scale $\mu_0=m_H/2$
 the \NNNLO{} correction is in fact very small (2.4\%) and,
 accordingly, the corresponding scheme spread in the right-hand plot of
 Fig.~\ref{fig:oldvsnewJVE} is very small. To investigate whether this
 is a general feature of the new scheme prescription, one can examine
 the uncertainty band at a different central scale. We have done this
 in App.~\ref{app:mho2vsmh}, where it shown that in that case the
 scheme variation contributes significantly to the size of the
 theoretical uncertainty.  

 An alternative way to address the issue of accidentally small scheme
 $(b)$ variation is to introduce a prescription that probes the scheme
 variation at different scales (where the size of the \NNNLO{}
 corrections may be more sizeable). For instance, one could determine
 the scheme uncertainty by adding to the usual envelope the spread
 between schemes $(a)$ and $(b)$ at different scales. 
 We therefore investigate the following $(b\!:\! a)$ prescription: to
 define the uncertainty band, we take the envelope of scheme $(a)$
 with its set of 7 scale variations (and $Q$ and $R_0$ variations) and
 additionally the 7 scale variations of
 \begin{equation}
   \label{eq:b:a-scalevar}
   \epsilon^{(a)}_{\mu_0, \mu_0} (\ptjv) + 
   \epsilon^{(b)}_{\mu_R, \mu_F} (\ptjv) -
   \epsilon^{(a)}_{\mu_R, \mu_F} (\ptjv)\,,
 \end{equation}
where we have included explicit subscript labels for the
renormalisation and factorisation scales.
By sampling Eq.~(\ref{eq:b:a-scalevar}) over 7 scale choices, one
explores the maximum difference between schemes $(a)$ and $(b)$ (with
identical scale choices for the two schemes) and applies that
difference as an additional uncertainty relative to the result of
scheme $(a)$ for its central scale choice $\mu_R = \mu_F = \mu_0$.
In this way one avoids the problem that the difference between schemes
$(b)$ and $(a)$ may be accidentally small for the central scale
choice.
This approach also avoids the potential risk of double counting of
uncertainties that would come were one simply to take the envelope of
schemes $(a)$ and $(b)$, each with 7 scale variations.

\begin{figure}
  \centering
  \includegraphics[width=0.49\textwidth]{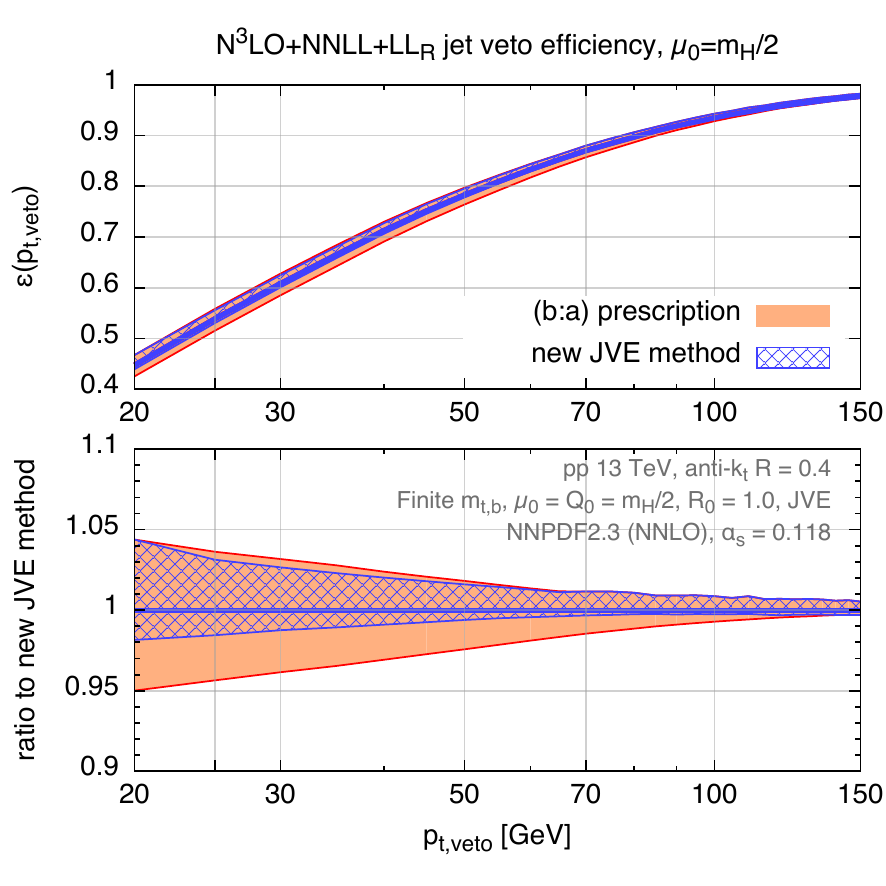}\hfill
  \includegraphics[width=0.49\textwidth]{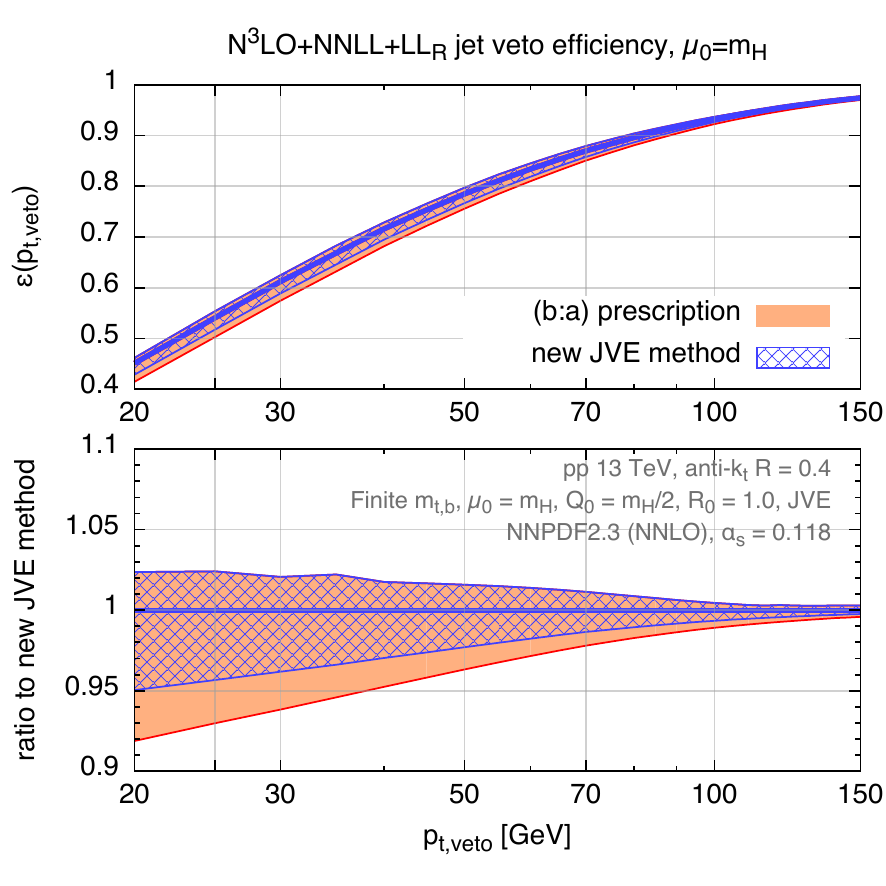}
  \caption{Jet-veto efficiency with uncertainty band obtained with our
    nominal JVE method and with the $(b\!:\!a)$ prescription as
    described in the text at central scales $\mu_0=m_H/2$ (left) and
    $\mu_0=m_H$ (right).}
  \label{fig:schemaacolb}
\end{figure}

 The comparison between the new JVE
 prescription to the $(b\!:\!a)$ procedure is shown in
 Fig~\ref{fig:schemaacolb}. We see that the $(b\!:\!a)$ prescription
 gives rise to only marginally larger uncertainties. Moreover, we
 have found that the $(b\!:\!a)$ prescription gives rise to enlarged
 uncertainties in the tail of the leading jet $p_t$ distribution. For
 these reasons, and due the fact that this procedure is more
 cumbersome and relies on a non-standard method to assess the error,
 we do not adopt it as our default prescription.

Besides the different efficiency scheme variation, another important
difference between our new JVE prescription and the original
one~\cite{Banfi:2012yh} is the range of variation for the resummation
scale $Q$.  Instead of varying it in the range $\{m_H/4,m_H\}$ as done
originally, we now restrict ourselves to the smaller range
$\{m_H/3,3/4\, m_H\}$. The main reason for this is that one wants the
one-jet cross section at large $p_t$ to be insensitive to the
resummation, therefore one should ensure that the resummation is
correctly turned off at large $\ptjv$ values. 
The scale at which the resummation is turned off is determined by the
resummation scale $Q$.
Making the choice $Q=m_H$ has the effect of starting the resummation
in a region of relatively high $p_t$, where the underlying soft and
collinear approximations are far from being valid.

\begin{figure}
  \centering
  \includegraphics[width=0.49\textwidth]{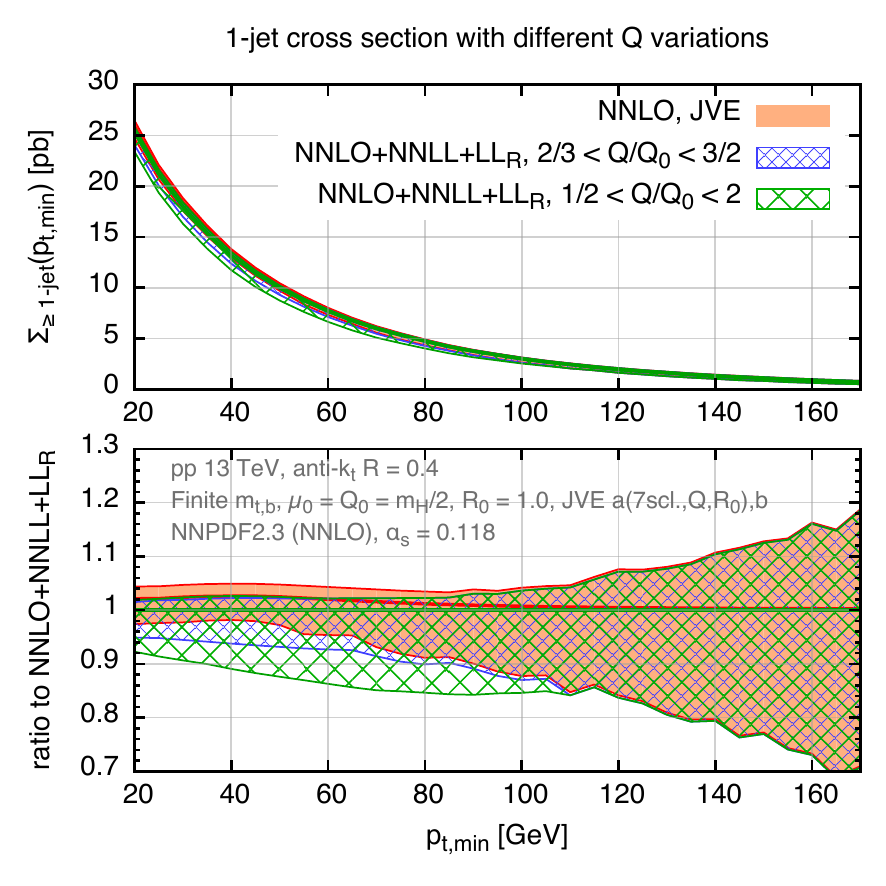}\hfill
  \includegraphics[width=0.49\textwidth]{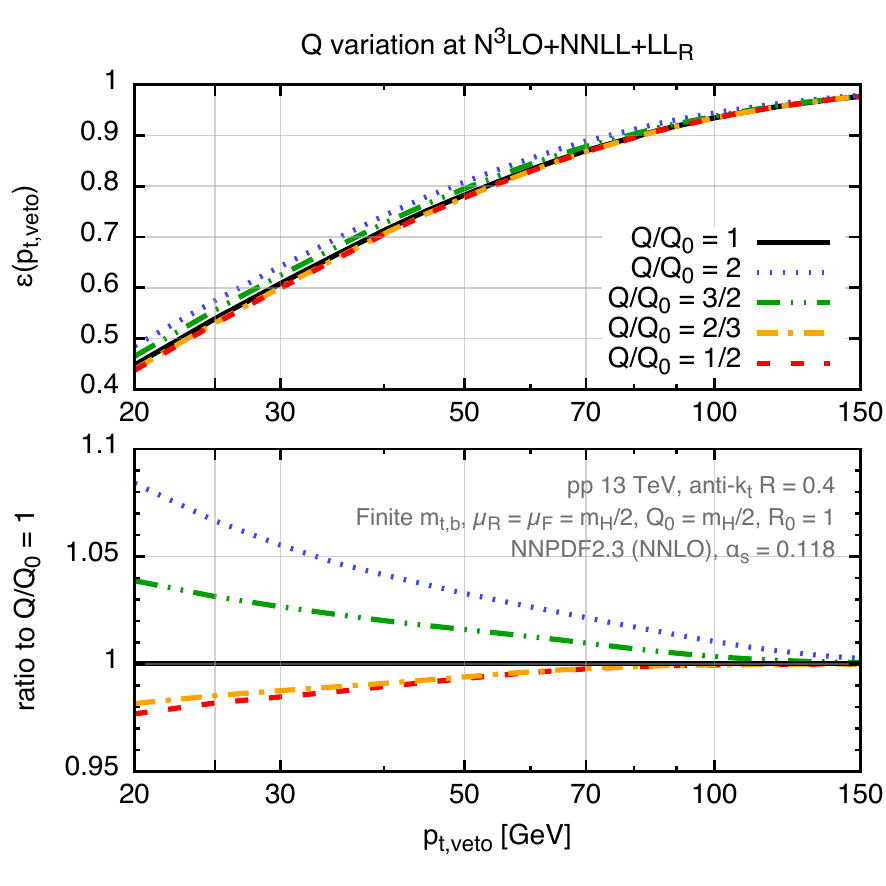}
  \caption{Left: one-jet cross section at \NNLO{} (red/solid band) and
    NNLO+NNLL+LL$_R$ (blue/hatched band), with uncertainty obtained
    with the new JVE method as described in the text, compared to the
    NNLO+NNLL+LL$_R$ with a $Q$ variation in the range $\{m_H/4,m_H\}$
    (green/hatched). Right: jet-veto efficiency at N$^3$LO+NNLL+LL$_R$
    for different values of the resummation scale $Q$, as used in the
    old ($Q/Q_0 = 1/2$, $Q/Q_0 = 2$) and new ($Q/Q_0 = 2/3$,
    $Q/Q_0 = 3/2$) formulations of the JVE method, where
    $Q_0 = m_H/2$.}
  \label{fig:Qvariation}
\end{figure}

In the left-hand plot of Fig.~\ref{fig:Qvariation} we show the one-jet
cross section at NNLO+NNLL+LL$_R$ with uncertainties obtained with the
new formulation of the JVE method both with a $Q$ variation range of
$\{m_H/4,m_H\}$ (green/hatched band) and $\{m_H/3,3/4\, m_H\}$
(blue/hatched).
The fixed order result (red/solid) is also shown for
comparison. We observe that while the effect of resummation on the
central value is very moderate, the band obtained with the old $Q$
variation range is substantially larger all the way up to
$p_t \sim 100\,\GeV$. 
The right-hand plot of Fig.~\ref{fig:Qvariation} shows the jet-veto
efficiency at N$^3$LO+NNLL+LL$_R$ for the values of the resummation
scale $Q$ used in the old ($Q = m_H/4$, $Q = m_H$) and in the new
($Q = m_H/3$, $Q = 3/4\,m_H$) prescription for the JVE method. While
the curve corresponding to the upper variation changes significantly
when reducing $Q$ from $m_H$ to $3/4\,m_H$, the curve corresponding to
the lower edge is largely unaffected by the change in the variation
range.
The insensitivity to the choice of the lower end of the range for $Q$
motivates a simple symmetric choice for the resummation scale range.

\begin{figure}
  \centering
  \includegraphics[width=0.49\textwidth]{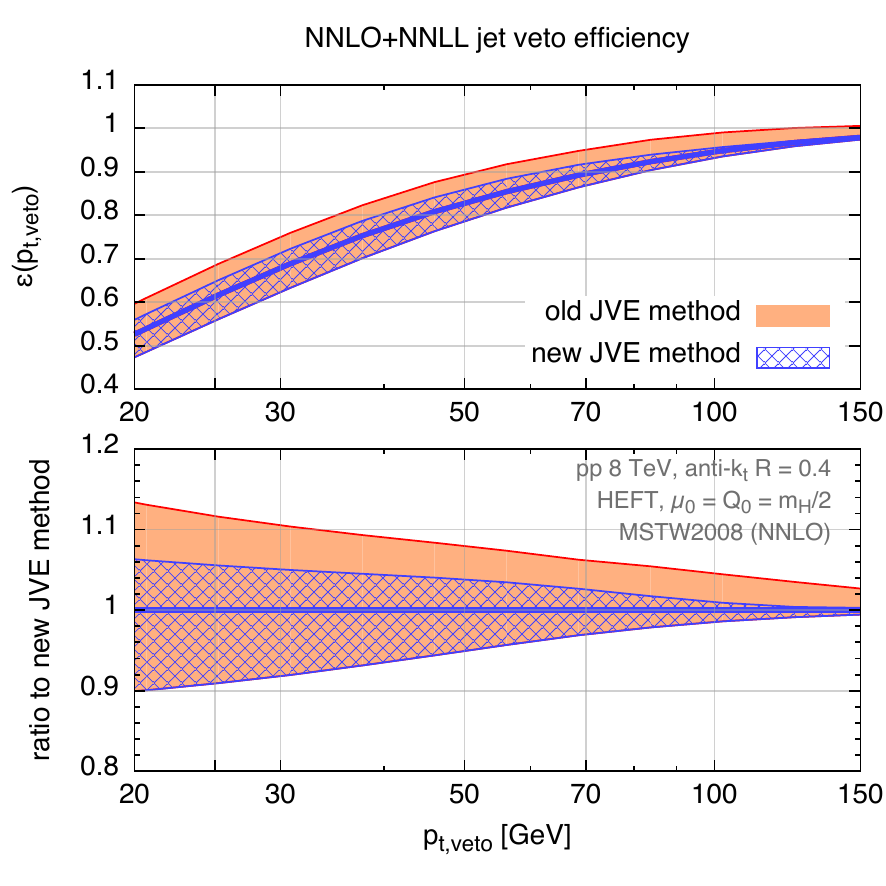}
  \caption{Jet-veto efficiency at NNLO+NNLL at 8\,TeV with uncertainty
    bands obtained with the original formulation of the JVE
    method~\cite{Banfi:2012jm} (red/solid) compared to the prediction
    obtained with the new JVE method as defined in the text
    (blue/hatched).}
  \label{fig:oldvsnew8TeV-full-prediction}
\end{figure}

Finally, it is interesting to see how the original prediction of
ref.~\cite{Banfi:2012jm} changes with the new prescription for the JVE
uncertainty. Fig.~\ref{fig:oldvsnew8TeV-full-prediction} shows the
comparison between the old and new JVE methods for the NNLO+NNLL
efficiency at 8\,TeV. We observe that the new prescription leads to a
reduction of the upper part of the uncertainty band. 
At low values of $\ptjv$ this reduction is mainly driven by the
reduction in the $Q$ variation range (cf.~Fig.~\ref{fig:Qvariation}),
while at large $\ptjv$ scheme $(c)$ gives a significant contribution
to the theoretical uncertainty (cf.\ Fig.~\ref{fig:oldJVE8vs13}).
To conclude this section, we remark that when the original formulation
of the JVE method was proposed, the NNLO corrections showed a somewhat
problematic convergence, therefore a more conservative approach to the
uncertainty estimate seemed appropriate. Now that the computation of
the N$^3$LO correction shows a much better convergence of the
perturbative series, extensive study has led us to believe that the
new formulation of the JVE method is more appropriate.

\section{Choice of the central scale}
\label{app:mho2vsmh}

Results presented in the main text are obtained using $m_H/2$ as a
central scale choice. 
This choice, rather than $m_H$, is motivated by the better convergence
of the perturbative expansion and by the fact that soft emissions and
virtual corrections that contribute substantially to the cross-section
tend to have scales that are typically lower than $m_H$.
It is similar also to the choice of $H_T/2$ or $p_{t,\text{jet}}$ that
is often used in processes with more complex final states.
Nevertheless it is interesting to examine how much our central results
and the uncertainties change when $m_H$ is adopted as a central scale.

Table~\ref{tab:sigmatotmh} shows the input numerical values for the total
and one-jet cross section, with and without mass effects, up to
${\cal O}(\alpha_s^3)$, with uncertainties obtained through scale
variation and using NNLO PDFs and $\alpha_s$ using a central scale
$m_H$. These numbers are to be compared to those at scale $m_H/2$,
Table~\ref{tab:sigmatot}.

\begin{table}
\begin{center}
{\renewcommand{\arraystretch}{1.1}
  \begin{tabular}{c||c|c||c|c||c|c}
LHC 13 TeV [pb]& $\sigmatot{2}$ &  $\sigmatot{3}$ & $\sigma_{\rm 1j \ge 25 {\rm GeV}}^{\rm NLO}$ &  $\sigma_{\rm 1j \ge 25 {\rm GeV}}^{\rm NNLO}$ & $\sigma_{\rm 1j \ge 30 {\rm GeV}}^{\rm NLO}$ &  $\sigma_{\rm 1j \ge 30 {\rm GeV}}^{\rm NNLO}$ 
    \\[0.2em] \hline 
EFT      & $41.1^{+4.4}_{-4.3}$ & $44.8^{+1.3}_{-2.5}$ & $16.9^{+3.5}_{-2.9}$ & $20.2^{+1.4}_{-2.0}$ & $14.4^{+3.0}_{-2.5}$ & $17.1^{+1.2}_{-1.6}$ \\  
$t$-only & $42.9^{+4.7}_{-4.5}$ & $46.6^{+1.6}_{-2.7}$ & $17.3^{+3.5}_{-3.0}$ & $20.6^{+1.4}_{-2.0}$ & $14.6^{+3.0}_{-2.5}$ & $17.4^{+1.2}_{-1.6}$ \\
$t,b$    & $40.8^{+4.6}_{-4.3}$ & $44.5^{+1.5}_{-2.5}$ & $17.1^{+3.5}_{-3.0}$ & $20.5^{+1.4}_{-2.0}$ & $14.6^{+3.0}_{-2.5}$ & $17.4^{+1.2}_{-1.6}$ 
  \end{tabular}}
\end{center}
 \caption{Total cross section at NNLO ($\sigmatot{2}$) and
   at \NNNLO{} ($\sigmatot{3}$), and the one-jet
   cross-section $\sigma_{\rm 1j}$ at NLO and NNLO for central scale
   $\mu_0 = m_H$, with and without mass effects as explained in the
   text. Uncertainties are obtained with a 7-point renormalisation and
   factorisation scale variation.}
 \label{tab:sigmatotmh} 
\end{table}

\begin{figure}[htp]
  \centering
  \includegraphics[width=0.49\textwidth]{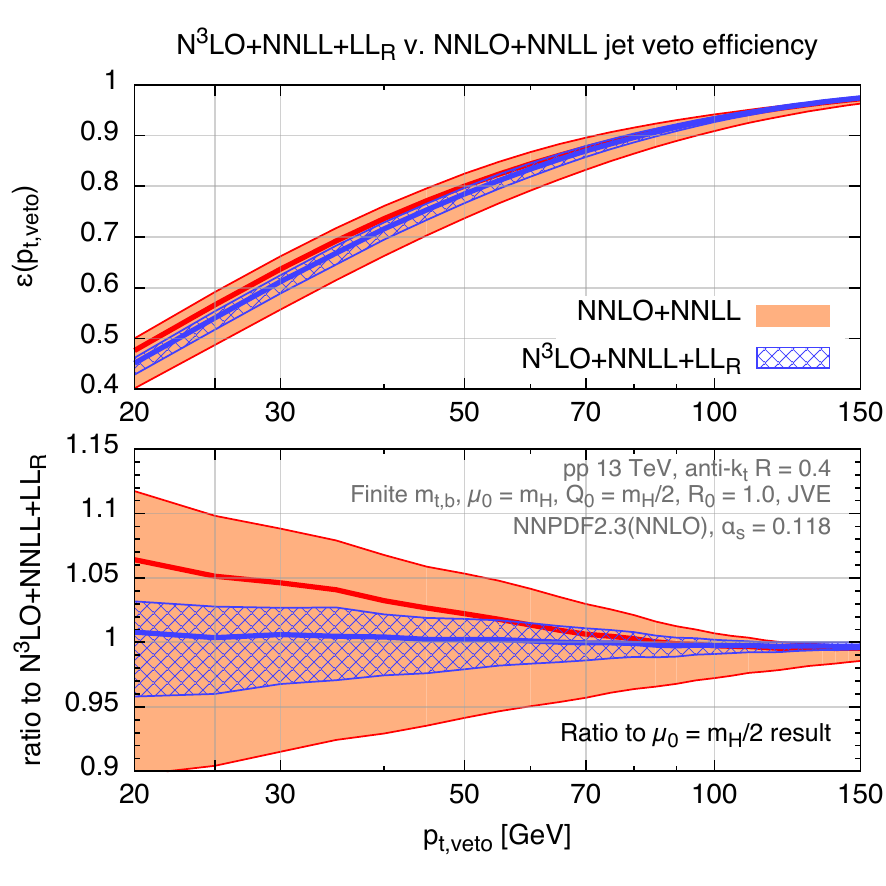}\hfill
  \includegraphics[width=0.49\textwidth]{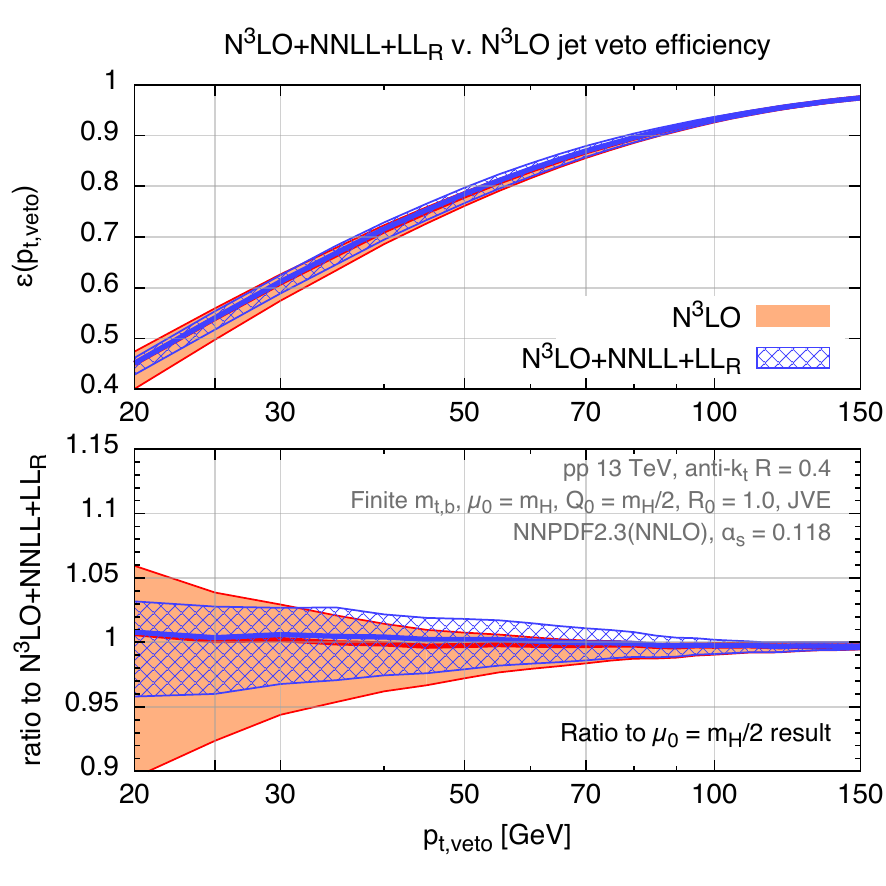}
  \caption{N$^3$LO+NNLL+LL$_R$ best prediction for the jet-veto efficiency (blue/hatched)
    compared to NNLO+NNLL (left) and fixed-order at N$^3$LO
    (right) at $\mu_0=m_H$. The lower panel shows the ratio to the $\mu_0=m_H/2$ result.
  }
  \label{fig:bestprediction-efficiency-mh}
\end{figure}

\begin{figure}[htp]
  \centering
  \includegraphics[width=0.49\textwidth]{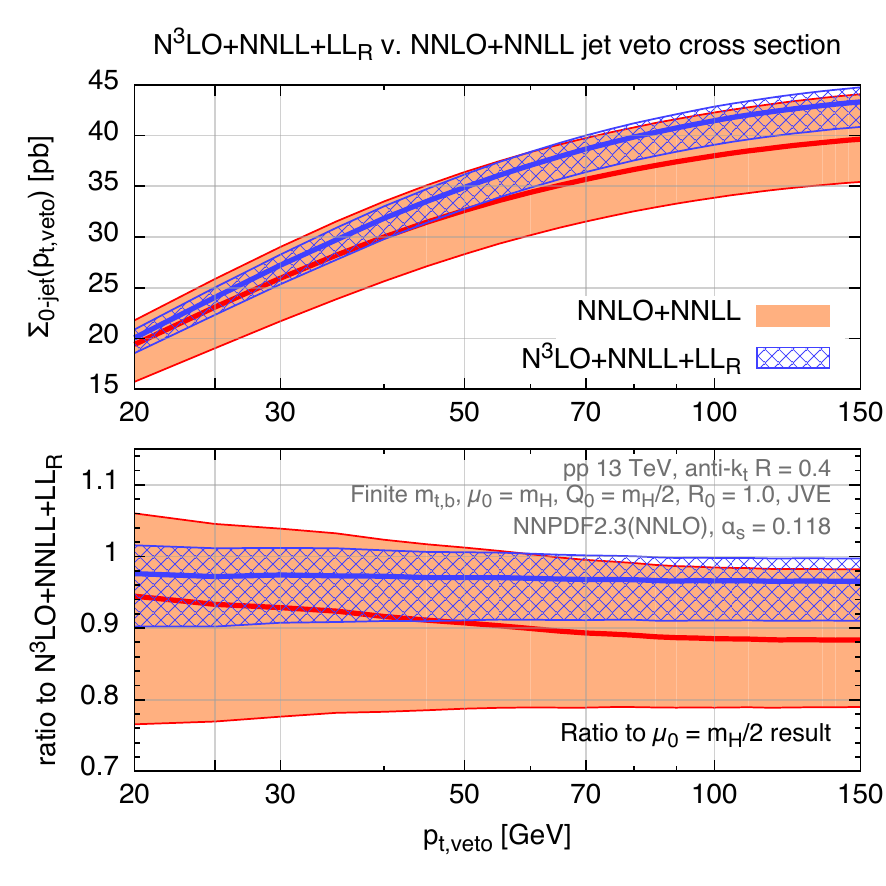}\hfill
  \includegraphics[width=0.49\textwidth]{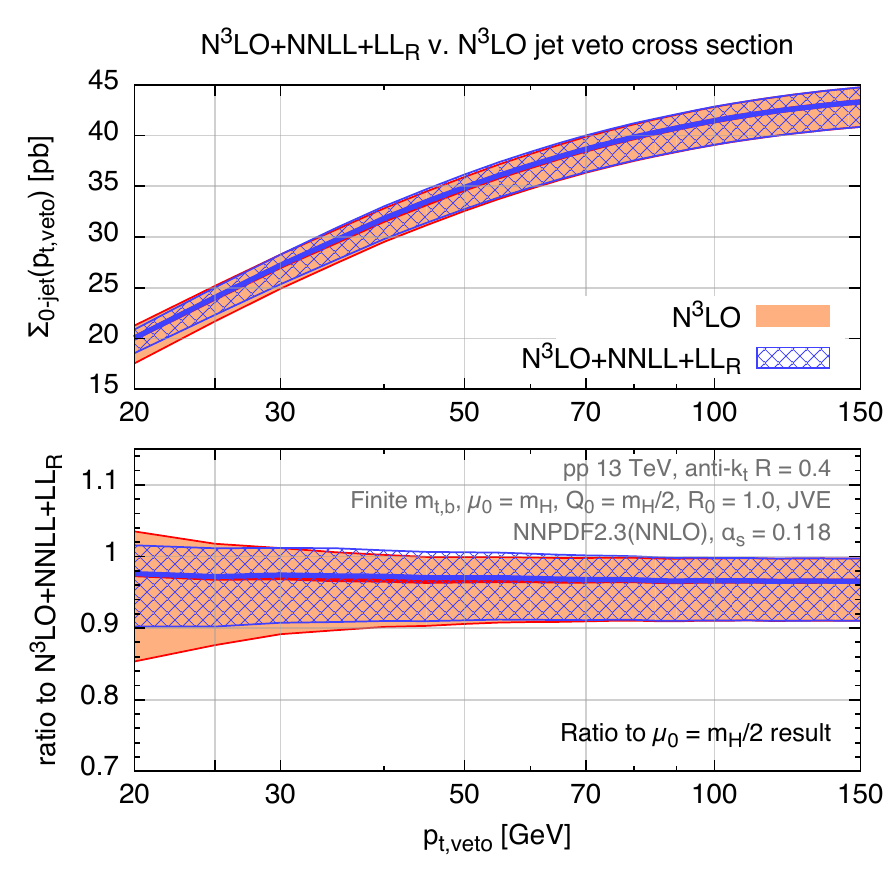}
  \caption{N$^3$LO+NNLL+LL$_R$ best prediction for the jet-veto cross section (blue/hatched)
    compared to NNLO+NNLL (left) and fixed-order at N$^3$LO
    (right) at $\mu_0=m_H$. The lower panel shows the ratio to the $\mu_0=m_H/2$ result.
  }
  \label{fig:bestprediction-sigma-mh}
\end{figure}

In Fig.~\ref{fig:bestprediction-efficiency-mh} we show a comparison of
the $\NNNLO+\NNLL+\LLR$ results to $\NNLO+\NNLL$ results (left) and to
$\NNNLO$ (right) at central scale $m_H$. This figure is to be compared
to the similar one at central scale $m_H/2$,
Fig.~\ref{fig:bestprediction-efficiency}. It is clear that at scale
$m_H$ uncertainties are somehow larger, this is particularly the case
for the \NNNLO{} prediction. Accordingly, uncertainty bands overlap
slightly better at scale $m_H$. Still, the change in the central value
at $\NNNLO+\NNLL+\LLR$ is very small when using $m_H$ rather than
$m_H/2$. The corresponding plots for the 0-jet cross section are shown in 
Fig.~\ref{fig:bestprediction-sigma-mh}.
Results for
the efficiency and 0-jet cross section are reported in Tab.~\ref{tab:13-TeV-0jet-mh}.

\begin{table}
\begin{center}
{\renewcommand{\arraystretch}{1.1}
  \begin{tabular}{c|c|c|c|c}
    LHC 13 TeV
    & $\epsilon^{{\rm N^3LO+NNLL+LL_R}}$ & \,$\Sigma^{{\rm
                                N^3LO+NNLL+LL_R}}_\text{0-jet}\,{\rm [pb]}$\,  &
                                                          \,$\Sigma^{{\rm
                                                          N^3LO}}_\text{0-jet}$ &
                                                          \,$\Sigma^{{\rm
                                                          NNLO+NNLL}}_\text{0-jet}\,$
    \\[0.2em] \hline 
    $\ptjv=25\,{\rm GeV}$ & $0.541^{+0.013}_{-0.023}$ & $24.0^{+1.0}_{-1.7}$ & $24.0^{+1.2}_{-2.3}$ & $23.1^{+2.8}_{-4.0}$ \\
    $\ptjv=30\,{\rm GeV}$ & $0.612^{+0.013}_{-0.023}$ & $27.2^{+1.1}_{-1.9}$ &  $27.1^{+1.2}_{-2.2}$ & $25.9^{+3.1}_{-4.2}$        
  \end{tabular}}
\end{center}
 \caption{Predictions for the jet-veto efficiency and cross
   section at N$^3$LO+NNLL+LL$_{R}$, compared to the N$^3$LO and NNLO+NNLL cross
   sections.
   The uncertainty in the fixed-order prediction is obtained using the
   JVE method. All numbers include the effect of top and bottom quark
   masses, treated as described in the text, and are for a central scale $\mu_0=m_H$.
}
 \label{tab:13-TeV-0jet-mh} 
\end{table}

\begin{figure}[htbp]
  \centering
  \includegraphics[width=0.49\textwidth]{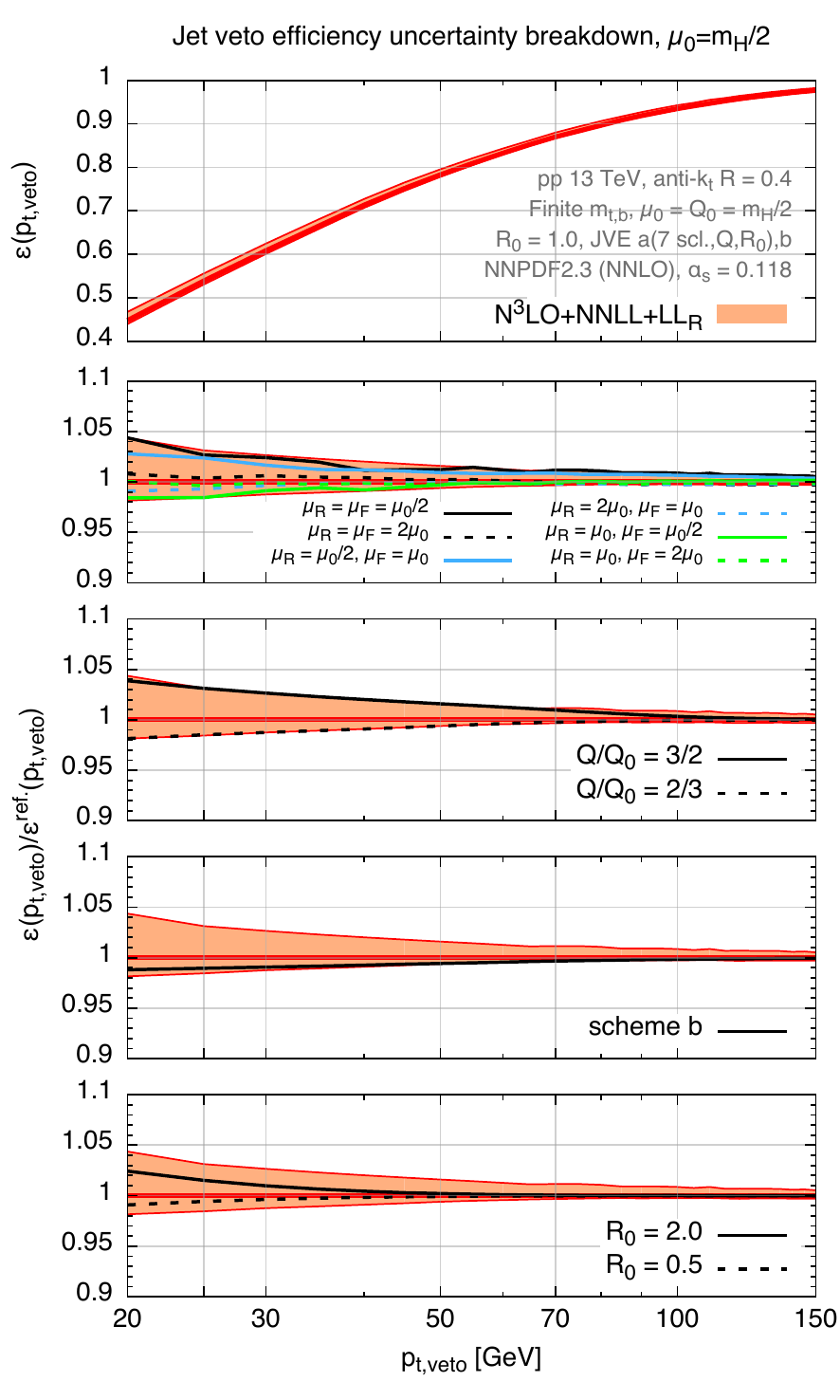}\hfill
  \includegraphics[width=0.49\textwidth]{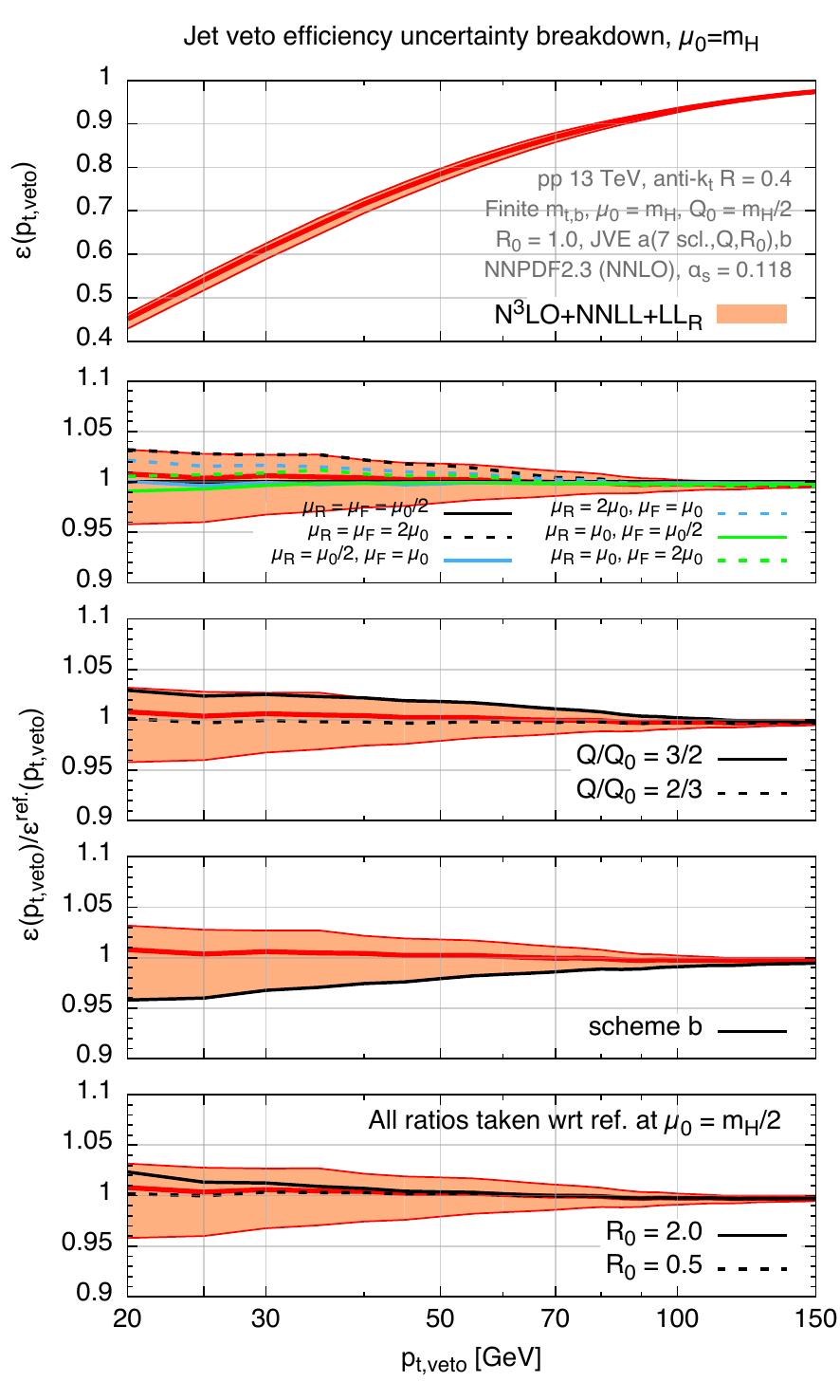}
  \caption{N$^3$LO+NNLL+LL$_R$ best prediction for the jet-veto
    efficiency (red band), with a breakdown (lower panels) comparing the
    overall relative uncertainty envelope to the different contributions
    from which it is built up. 
    The left and right-hand plots show results respectively for scale
    choices $\mu_0 = m_H/2$ and $m_H$.
    In both plots, ratios are taken with respect to a reference result
    determined with $\mu_0 = m_H/2$.
  }
  \label{fig:uncertainty-detail}
\end{figure}

To gain insight into the differences between the two scale choices,
Fig.~\ref{fig:uncertainty-detail} shows the breakdown into different
sources of uncertainty using $m_H/2$ (left) and $m_H$ (right) as a
central scale choice. 
We notice that for the
central scale $m_H/2$ the full uncertainty band is determined by the
scale variations (both $Q$ and $\mu_R,\mu_F$), while scheme and $R_0$
variation give rise to a lower uncertainty. For central scale $m_H$
the upper edge of the band is still set by scale variation, while the
lower one is determined by the scheme variation, and $R_0$ variation
has still no impact on the final uncertainty band.
The difference in the impact of the scheme variation at the two
different scales is a consequence two facts: (a) at scale $m_H/2$
the \NNNLO{} correction is only a 2\% correction, while it amounts to
9\% at scale $m_H$; and (b) in our updated JVE approach, the scheme-variation is now
sensitive only to the ambiguity of including (or not) the \NNNLO{}
correction to the total cross section in the efficiency. 

\begin{figure}
  \centering
  \includegraphics[width=0.49\textwidth]{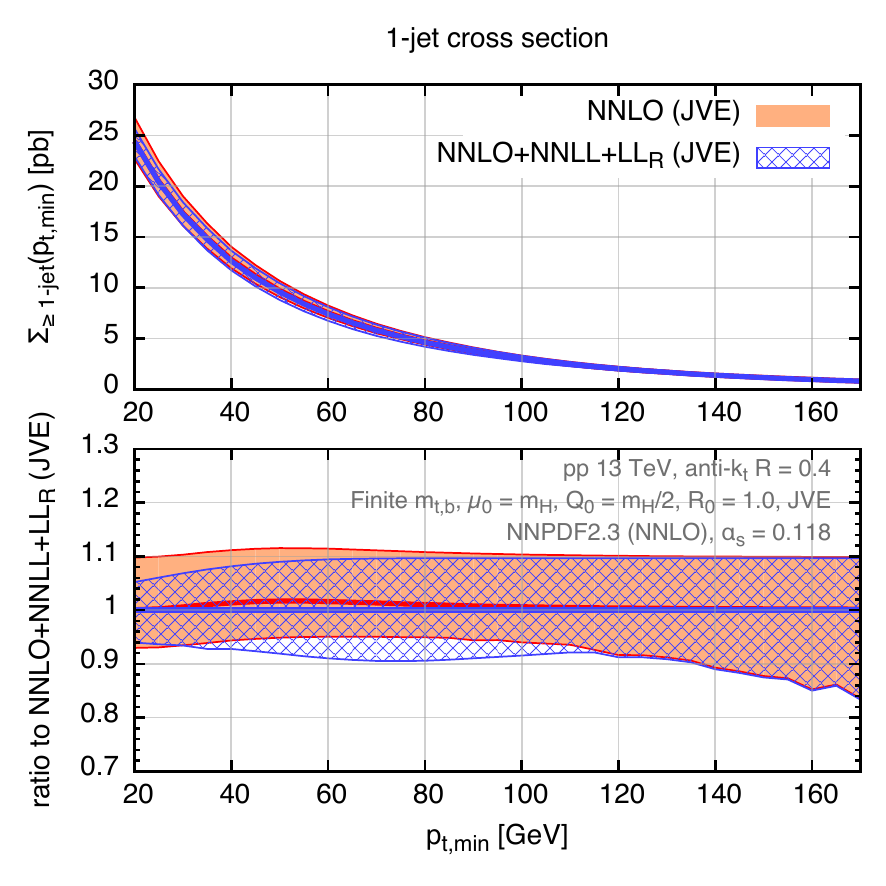}\hfill
  \includegraphics[width=0.49\textwidth]{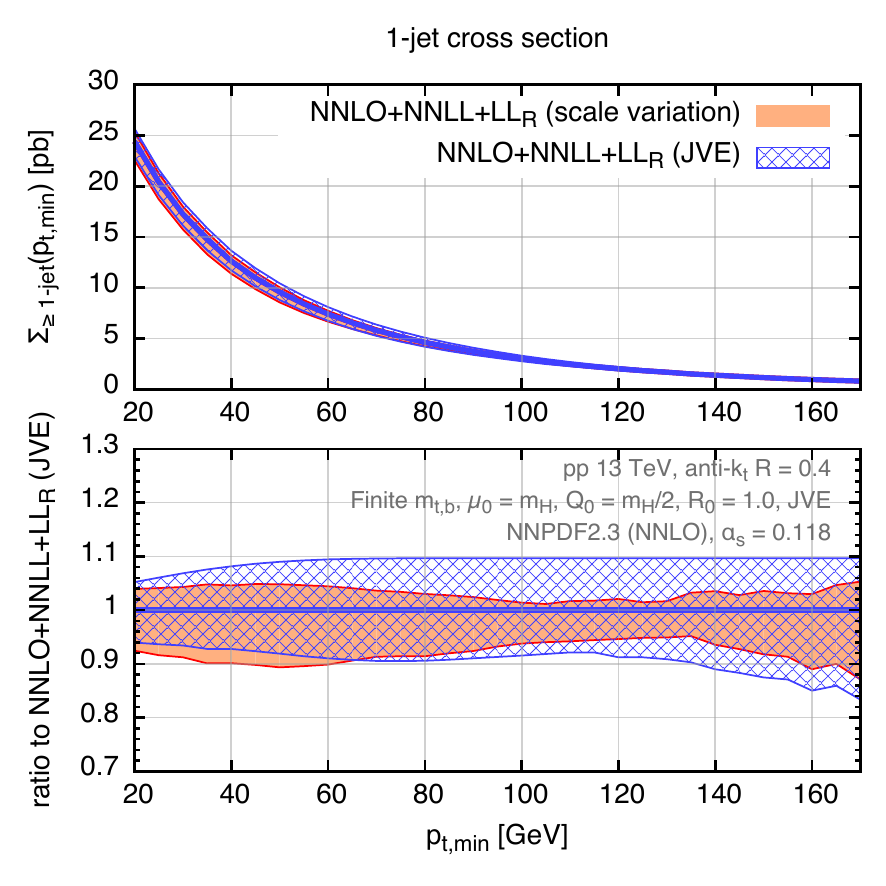}
  \caption{Best prediction for the inclusive one-jet cross section
    (blue/hatched) compared to fixed-order at NNLO (left) and to the
    matched result with direct scale variation for the uncertainty as
    explained in the text (right). The central renormalisation and
    factorisation scales are set to $\mu_0=m_H$. The lower panel shows the
    ratio to the central value at $\mu_0=m_H$.}
  \label{fig:1jetxsatmh}
\end{figure}

\begin{table}
  \begin{center}
    {\renewcommand{\arraystretch}{1.1}
      \begin{tabular}{c|c|c}
        LHC 13 TeV
        &\, $\Sigma^{{\rm
          NNLO+NNLL+LL_R}}_{\ge \text{1-jet}}\,{\rm [pb]}$\,  &
                                                               \,$\Sigma^{{\rm
                                                               NNLO}}_{\ge\text{
                                                               1-jet}}\,{\rm
                                                               [pb]}$
        \\[0.2em]\hline
    $p_{\rm t, min}=25\,{\rm GeV}$ & $20.4^{+1.2}_{-1.3}$ & $20.5^{+2.0}_{-1.5}$\\
    $p_{\rm t,min}=30\,{\rm GeV}$ & $17.2^{+1.2}_{-1.1}$ & $17.3^{+1.7}_{-1.2}$
  \end{tabular}}
\end{center}
 \caption{Predictions for the inclusive one-jet cross section at
   NNLO+NNLL+LL$_{R}$ and NNLO. The uncertainty in the fixed-order
   prediction is obtained using the JVE method. All numbers include
   the effect of top and bottom quark masses, treated as described in
   the text, and are for a central scale $\mu_0=m_H$.
}
 \label{tab:13-TeV-1jet-mh} 
\end{table}

Next, in Fig.~\ref{fig:1jetxsatmh} we show the inclusive one-jet cross
section (blue/hatched) compared to fixed-order at NNLO (left) and to
the matched result with direct scale variation for the uncertainty as
explained in the text (right) at central scale $m_H$. 
Corresponding numerical values for the one-jet cross section are
reported in Tab.~\ref{tab:13-TeV-1jet-mh}.
From the right-hand plot of Fig.~\ref{fig:1jetxsatmh}, one notices
that the JVE uncertainty band, especially its upper edge, is larger
than scale variation even at transverse momenta of the order of $m_H$.
This larger uncertainty for the JVE result appears to be associated
with the variation between schemes $(a)$ and $(b)$, which differ by
about $10\%$ over a range of $p_{t,\min}$, a consequence of the nearly
$10\%$ difference between $\sigma_\text{tot,2}$ and
$\sigma_\text{tot,3}$ that is visible in table~\ref{tab:sigmatotmh}.
This effect is not present for the results with central scale $\mu_0
= m_H/2$, Fig.~\ref{fig:1jetxs}, where the difference between the two
schemes is much smaller.
However, for large values of $p_{t,\min}$ the uncertainty on the
$\mu_0 = m_H/2$ results grows more rapidly, perhaps a consequence of
the fact such a scale choice is not appropriate at high $p_t$.


\section{Small-$R$ correction factor}
\label{app:smallr}

In Ref.~\cite{Dasgupta:2014yra}, small-$R$ effects for jet vetoes were
resummed through the introduction of a ``fragmentation'' function
$f^\text{hardest}(z,t)$ for the distribution of the momentum fraction
$z$ carried by the hardest subjet resulting from the fragmentation of a gluon.
The quantity ${\cal Z}(t)$ used in Eq.~(\ref{eq:fcorrelmod}) is the
first logarithmic moment of this fragmentation function,
\begin{align}
  \label{eq:lnz}
  {\cal Z}(t) & \equiv   \int_0^1 dz f^\text{hardest}(z,t) \ln z \nonumber
  \\
  & \simeq t\bigg[
      \frac{1}{72} C_A \left(131 - 12 \pi ^2 - 132\ln 2\right)
      + \frac{1}{36} n_f T_R (-23+24 \ln 2)
  \bigg]\nonumber
  \\
  & + \frac{t^2}{2!} \big(  0.206672 C_A^2 
      + 0.771751 C_A n_f T_R \nonumber
      \\
      & \hspace{25mm} - 0.739641 C_F n_f T_R 
      + 0.117861 n_f^2 T_R^2 \big)\nonumber
  \\
  & + \frac{t^3}{3!} \big(-0.20228(4) C_A^3 
      - 0.53612(2) C_A^2 n_f T_R 
      - 0.062679(8) C_A C_F n_f T_R \nonumber
      \\
      & \hspace{25mm} + 0.54199(2) C_F^2 n_f T_R
      - 0.577215(3) C_A n_f^2 T_R^2 \nonumber
      \\
      & \hspace{25mm} + 0.431055(4) C_F n_f^2 T_R^2 
      - 0.0785743(5) n_f^3 T_R^3 \big) \nonumber
      \\
   & + \frac{t^4}{4!} c_4^\textrm{fit}
   + \frac{t^5}{5!} c_5^\textrm{fit}
   + \frac{t^6}{6!} c_6^\textrm{fit}
   + \frac{t^7}{7!} c_7^\textrm{fit}
   + \frac{t^8}{8!} c_8^\textrm{fit}\,,
\end{align}
where the coefficients up to $t^3$ are the actual terms of the full
Taylor expansion of ${\cal Z}(t)$, while those from $t^4$ to $t^8$,
given in table~\ref{tab:lnz-fit}, are chosen so as to provide a good
fit to the full numerical form for ${\cal Z}(t)$ as calculated in
Ref.~\cite{Dasgupta:2014yra}.
Coefficients are tabulated both for $n_f=4$ and $n_f=5$.%
\footnote{The results of this paper use the $n_f=5$ values.}
As such, these higher order coefficients are not the actual values of
the coefficients of the Taylor series for ${\cal Z}(t)$, since yet
higher-order contributions might be partly absorbed in the fit.
Eq.~(\ref{eq:lnz}) reproduces the full all-order result at the
per mil level in the range $0<t<1$, which should be more than
adequate for phenomenological applications.

\begin{table}
  \centering
  \begin{tabular}{cccccc}
    \toprule
     & $c_4^\textrm{fit}$ & $c_5^\textrm{fit}$ & $c_6^\textrm{fit}$ 
     & $c_7^\textrm{fit}$ & $c_8^\textrm{fit}$
    \\
    \midrule
     $n_f=5$ & 133.55 & -478.55 & -1226.87 & 22549.99 & -77020.08 \\[4pt]
     $n_f=4$ & 100.69 & -352.10 &  -858.44 & 15819.97 & -53597.50 \\[2pt]
    \bottomrule
  \end{tabular}
  \caption{
    Results of a fit to parametrise the all-order result of the integral in
    Eq.~(\ref{eq:lnz}). 
    Values are given for $n_f=4$ and $n_f=5$. 
    The fitted curve is accurate to 0.1\% in the $t\in[0,1]$ range.
  }
  \label{tab:lnz-fit}
\end{table}

For the purposes of matching, it is useful to have the $\as$ expansion
of $\mathcal{F}^\text{correl}_{\LLR}(R)$ up to $\as^3$.
The $\as^2 L$ and $\as^3 L^2$  terms are known from previous work.
Once one includes \LLR resummation there is an additional $\as^3 L
\ln^2 R$ term, which 
receives contributions from both the
order $t$ and $t^2$ terms in Eq.~(\ref{eq:lnz}), because $t$ itself
has an all-order expansion in powers of  $\as \ln R$.
It is given by
\begin{multline}
  \label{eq:h31}
  \mathcal{F}^\text{correl}_{\LLR,31}(R) = 
  \left(\frac{\as}{2\pi}\right)^3 L \cdot 
  16C_A \ln^2\frac{R}{R_0}\Big[ 
        1.803136 C_A^2 - 0.589237 n_f 2 T_R C_A 
        \\
        + 0.36982 C_F n_f 2 T_R - 0.05893 n_f^2 4 T_R^2 \Big]\,.
\end{multline}


\end{document}